\documentclass[sigplan,10pt]{acmart}
\renewcommand\footnotetextcopyrightpermission[1]{}
\usepackage{tikz}
\usepackage{amsmath}
\usepackage{booktabs}
\usepackage{multirow}
\usepackage{xspace}
\usepackage{tabularx}
\usepackage{subfig}
\usepackage{listings}
\usepackage[]{xcolor}
\usepackage{paralist}
\usepackage[font=small,labelfont=bf]{caption}
\newcolumntype{?}{!{\vrule width 1pt}}

\begin{document}

\newcommand{\sys}{GSPMD\xspace}
\newcommand{\todo}[1]{{\color{red} TODO: #1}}

\title{\sys: General and Scalable Parallelization for ML Computation Graphs}


\author{Yuanzhong Xu, HyoukJoong Lee, Dehao Chen, Blake Hechtman, Yanping Huang, Rahul Joshi, Maxim Krikun, Dmitry Lepikhin, Andy Ly, Marcello Maggioni, Ruoming Pang, Noam Shazeer, Shibo Wang, Tao Wang, Yonghui Wu, Zhifeng Chen}
\affiliation{Google}

\begin{abstract}
We present \sys, an automatic, compiler-based parallelization system for
common machine learning
computations. It allows users to write programs in the same way as for a single
device, then give hints through a few annotations on how to distribute tensors,
based on which \sys will parallelize the computation. Its representation of
partitioning is simple yet general, allowing it to express different or
mixed paradigms of parallelism on a wide variety of models.


\sys infers the partitioning for every operator 
based on limited user
annotations, making it convenient to scale existing single-device programs. It
solves several technical challenges for production usage,
allowing \sys to achieve 50\% to 62\% compute utilization
on up to 2048 Cloud TPUv3 cores for models with up to one trillion parameters.

\end{abstract}

\settopmatter{printfolios=true}
\maketitle
\pagestyle{plain}




\section{Introduction}\label{sec:intro}

Recent development of neural networks has shown dramatic benefit from model scaling, 
creating a demand to parallelize computation in terms of both training data and
model parameters. Parallelism may be introduced in several ways:
data parallelism~\cite{krizhevsky2012imagenet} partitions training data, pipeline
parallelism~\cite{gpipe,pipedream,terapipe} partitions the computation graph, and within-layer
model parallelism~\cite{shoeybi2019megatron} partitions the weight and computation of each model layer.

We present \sys, 
a system that uses simple \emph{tensor sharding annotations} to achieve
different parallelism paradigms in a unified way, including data parallelism,
in-layer model parallelism, and novel strategies like image
spatial partitioning~\cite{spatial-partitioning} and weight-update/optimizer-state
sharding~\cite{xu2020automatic, rajbhandari2019zero}.
Although pipeline parallelism partitions the graph instead of individual operators/tensors,
\sys could still achieve it with the help of a simple wrapper library that reduces
pipelining to a tensor/operator partitioning problem.
\sys is flexible enough to express combinations of
these approaches, e.g., different layers could be partitioned in different manners,
and different approaches could be combined in the same layer.

\sys is generalized from the backend of GShard~\cite{gshard-arxive} based on our experiences
of model scaling beyond the mixture-of-expert (MoE) use case, and it has helped Google to scale
many deep learning models across several domains, including
language (e.g., LamBDA~\cite{lamda-blog}, GShard-M4~\cite{gshard-iclr}),
image (e.g., MetNet-2~\cite{espeholt2021skillful}),
and speech (e.g., BigSSL\cite{zhang2021bigssl}).
\sys as a shared, robust mechanism for different parallelism patterns is particularly relevant
moving forward because the ML community is increasingly investing into multimodality, where text,
image and audio are combined into a single model~\cite{dalle}.

\sys separates the concerns of machine learning model programming and parallelism.
It allows users to write programs with giant
tensors as if there were a single giant device. Then the user can insert
annotations in a few places that specify how tensors are distributed across
devices; \sys will run compiler passes that
complete the sharding specification on the entire computation graph, and transform it
into a mathematically equivalent, parallelized computation to run on each device. It
allows the users to focus on model building instead of sharding implementation, and
enables easy porting of existing single-device programs
to run at a much larger scale. To experiment with different partitioning strategies,
only the annotations need to be reconfigured.

\sys addresses several practical issues when applying automatic partitioning to production
models:
\begin{asparaitem}
\item Generating one program for each partition would increase compilation time significantly, 
so \sys instead produces a single program for all partitions.
This property is called Single Program Multiple Data (SPMD), and is crucial for scaling to thousands of partitions.

\item \sys supports unevenly partitioned dimensions, allowing any tensor to
be partitioned on arbitrary device meshes.
It is often a practical constraint for accelerators to require statically known shapes at compile time
in order to ease development.
Despite supporting uneven partitions, \sys is compatible with such constraints.

\item We implemented \sys as an extension to our production ML compiler, XLA~\cite{xla}. 
The implementation covers the full
set of operators in XLA, including those with complicated semantics like \texttt{Convolution}~\cite{xla-semantics}.
XLA is a unifying abstraction for multiple frameworks (TensorFlow~\cite{tensorflow}, Jax~\cite{jax2018github}, Pytorch~\cite{paszke2019pytorch} and Julia~\cite{bezanson2017julia}) and hardware platforms (CPUs, GPUs and Cloud TPUs~\cite{tpu}), making \sys reusable.


\item \sys supports nested patterns of parallelism; at per-operator level,
that means different types of dimensions could be partitioned across orthogonal subgroups of
devices. We have developed a recursive method for such nested patterns,
maximizing the generality of \sys without requiring excessive handwritten partitioning rules.
\end{asparaitem}

We demonstrate the capability of \sys by applying it on several categories of ML model, and
measuring the performance and memory scaling of model training on thousands of Cloud TPUv3~\cite{tpu} devices.
The use cases include image models, speech models, and sparse and dense language models.
By choosing intuitive
initial annotations, we can achieve close-to-linear memory and performance
scaling with respect to the number of devices.
\section{Background}\label{sec:background}

Modern ML models are typically defined as dataflow graphs of connected layers (subgraphs). Each layer has its model weights/parameters, and produces outputs that are referred to as ``activations''. Model training requires first computing the model's final output (forward pass), then computing the gradients of each layer weight (backward pass), which happens in the reverse layer order due to backward data dependencies. Model serving requires only the forward pass.


\subsection{Common parallelism patterns in ML workloads}
Below are a few typical parallelism patterns used in modern ML workloads.

\textbf{Data parallelism} is a technique for parallelizing training~\cite{krizhevsky2012imagenet}. Different devices (replicas) have the same copy of the model, but compute on different training data to produce local gradients. They collect and sum their gradients to keep in sync with each other. Such synchronization is typically implemented as an MPI-style \texttt{AllReduce} operator~\cite{mpi2.2}.

\textbf{Within-layer model parallelism} partitions the weight tensor of a layer across multiple devices~\cite{shoeybi2019megatron}. It may also require communication across devices, e.g., \texttt{AllReduce} when the layer sums over a partitioned dimension. Compared to data parallelism, this technique can help building larger models by sharding the weights.

\textbf{Spatial partitioning} is a technique to shard image input data along spatial dimensions~\cite{spatial-partitioning}, which helps fitting large image data on devices with limited memory capacity.

\textbf{Weight-update sharding or optimizer-state sharding} is an enhancement to data parallelism where the computation to apply combined gradients onto weights is sharded across replica devices~\cite{xu2020automatic,rajbhandari2019zero}. It is an optimization especially for expensive optimizers like ADAM~\cite{adam}.

\textbf{Pipeline parallelism} partitions the training graph into multiple stages that run on different devices~\cite{gpipe} to help build large models. Each stage sends its result to its downstream stage in the forward pass, and to its upstream stage in the backward pass. Due to data dependencies between the forward and backward passes, devices can be idle during part of the training step, which is known as a ``bubble''. The overhead of bubbles can be amortized with larger global training batch size. Pipelining can also help build large models.

\sys is designed to be a general solution for all of the above types of parallelism. It natively supports \textbf{in-operator} parallelism, which includes everything above except pipelining. Although pipeline parallelism is not in-operator partitioning, it could be \textbf{reduced to an operator partitioning} problem with the help of a wrapper library over existing code (Section~\ref{sec:spmd-pipe}) in some common cases, and \sys could still be used to shard individual stages in combination with other pipelining implementations for unsupported cases.

More importantly, \sys can easily express \textbf{nested patterns} of the above techniques. Section~\ref{sec:eval} shows examples of combining them in the Transformer~\cite{transformer} model.

\subsection{XLA and TensorFlow}
The automatic partitioner in \sys is implemented as transformation passes in the XLA compiler~\cite{xla}.
XLA defines a backend-agnostic intermediate representation (IR) calld HLO, which models a computation as a dataflow graph where nodes are operators and edges are tensors. 

Multiple front-end frameworks including TensorFlow~\cite{tensorflow}, JAX~\cite{jax2018github}, PyTorch~\cite{paszke2019pytorch} and Julia~\cite{bezanson2017julia} already have lowering logic to transform their graph representation to XLA HLO graph, which can then be compiled into target executables if the accelerator has a compiler backend for XLA. XLA has a much smaller set of operators than front-end frameworks like TensorFlow. This reduces the burden of implementing a partitioner without harming generality.

In this paper we demonstrate the use cases with models written in TensorFlow. TensorFlow offers a Python API for defining and running ML models. A component called the \emph{TF2XLA bridge} transforms the TensorFlow model into an equivalent XLA graph, so that XLA can compile it into a device executable. \sys is integrated to JAX with a slightly different API, but it is mapped to the same XLA abstraction.
\section{Tensor Sharding and Auto Completion}\label{sec:design}
\sys defines an intuitive and general representation of tensor sharding.
Following the philosophy of separation of concern, \sys has two independent
compiler transformations: sharding completion and per-operator partitioning.

\subsection{Representation of tensor sharding}
\begin{figure}[t!]
  \includegraphics[width=0.46\textwidth]{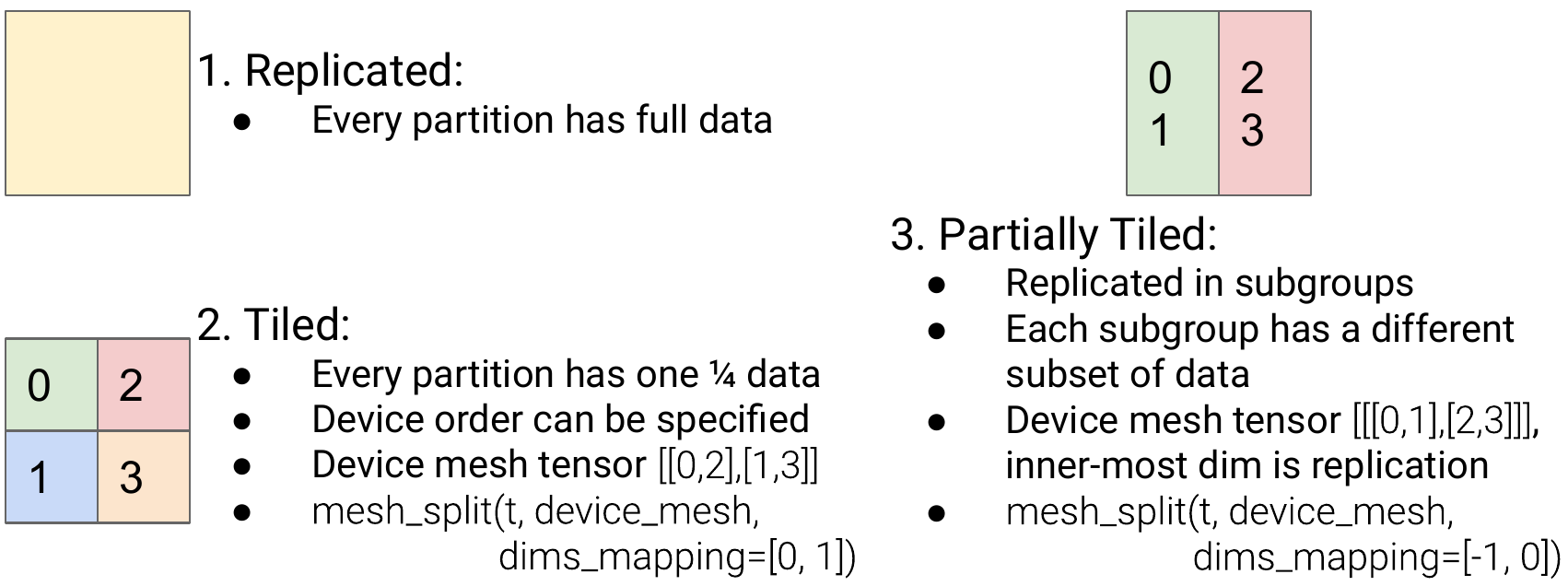}
\caption{Examples of the three types of sharding on a data tensor, and the \texttt{mesh\_split} API calls to represent them.}
\label{fig:sharding-rep}
\end{figure}

In \sys, each tensor will be assigned a \textbf{sharding} property, either
explicitly by the user as initial annotations, or by the sharding completion
pass. The sharding property specifies how the data
is distributed across devices. \sys defines three types of sharding
(see also Figure~\ref{fig:sharding-rep}).

\begin{asparaitem}
    \item Replicated. All devices have the same full data.
    \item Tiled. A tiled sharding contains a multi-dimensional tensor consisting of
    device IDs (e.g., [[0,2],[1,3]] in Figure~\ref{fig:sharding-rep}), which must have
    the same rank as the data tensor.
    Each data dimension is sharded across devices along the same dimension
    in the device tensor, and each device occupies the corresponding tile in the
    data tensor that matches its location in the device tensor. There is zero data
    duplication.
    \item Partially tiled (an extension to GShard~\cite{gshard-arxive}).
    The devices are first divided into equally sized subgroups,
    and the data tensor is replicated across devices in each subgroup but tiled across
    subgroups. Internally, it is also represented as a tensor of device IDs, with an
    additional dimension at the end for replication subgroups.
\end{asparaitem}

This representation is simple yet general, because it treats all tensor dimensions
in the same way and does not specialize for batch dimensions (data parallelism~\cite{krizhevsky2012imagenet}),
weight dimensions (model-parallelism or weight-update sharding~\cite{xu2020automatic}),
or image dimensions (spatial partitioning~\cite{spatial-partitioning}).

\sys provides a convenient abstraction on top of the above low-level sharding
representation. Devices can be organized in a logical multi-dimensional
tensor called a \textbf{device mesh}, which can  be used with an intuitive API.

\texttt{\bf mesh\_split(tensor, device\_mesh, dims\_mapping)} is the primary API
\sys provides for users. It generates a sharding annotation for \texttt{tensor},
based on the device mesh and a mapping from each data tensor dimension (\texttt{i})
to an optional device mesh dimension (\texttt{dims\_mapping[i]}). It uses -1 to
represent a non-existing mapping in \texttt{dims\_mapping}. Each device mesh
dimension should appear at most once in \texttt{dims\_mapping}. This simple API is
general enough to express all types of sharding: depending on whether
\texttt{dims\_mapping} contains all, part, or none of the mesh dimensions, it
can represent tiled, partially tiled, and replicated sharding.


When generating collective communication operators, \sys preserves the order of devices
in \texttt{device\_mesh}. Therefore, \texttt{device\_mesh} can be configured by the user
in order to optimize communication based on the topology of the device network. For
example, it could model the actual topology of device network, or reorder devices to
avoid long links.

In typical use cases, the user only needs to define the device mesh once, and then
focus on the mapping of tensor dimensions. However, \sys does not have any restrictions
if the user wants to use a different device mesh for each tensor. This
could be useful when different parts of the model have very different parallelism patterns.

\subsection{Examples of expressing in-operator parallelism}\label{sec:api-examples}
We explain a useful and succinct operator in TensorFlow
(and other libraries like numpy), the \emph{Einstein Summation} or
\texttt{Einsum}. The equivalent operator in XLA is \texttt{Dot}. It is a generalized
matrix multiply, where the user can define arbitrary numbers of dimensions of different types.
An \texttt{Einsum} can be expressed as a string equation, e.g.,
$ABC,ACD\to ABD$,
where $A$ is an embarrassingly parallel \emph{batch} dimension in both operands and the output,
\emph{C} is a \emph{contracting} dimension that only exist in the operands and will be sum-reduced,
while $B$ and $D$ are \emph{non-contracting} dimensions that exist in one operand and are inherited
by the output.

With \sys, the user can annotate the operands and the output to
combine different parallelism modes. For a typical fully connected projection layer, $BD,DF\to BF$,
the user can combine data- and model-parallelism by annotating
\begin{lstlisting}[basicstyle=\footnotesize\selectfont\ttfamily\linespread{0.6},escapeinside={&}{&},stepnumber=1,escapechar=\%]
    bd = mesh_split(bd, mesh, [0, -1])
    df = mesh_split(df, mesh, [-1, 1])
\end{lstlisting}
and \sys will auto-complete the output sharading with \texttt{mesh\_split(bf, mesh, [0,1])}
so that the input and output are partitioned on the batch dimension (data-parallelism)
across mesh dimension 0, while the layer weight \texttt{df} and the output are partitioned
on the feature dimension F (model-parallelism) on mesh dimension 1.

If the user further partitions the weights along the other mesh dimension, i.e., 
\begin{lstlisting}[basicstyle=\footnotesize\selectfont\ttfamily\linespread{0.6},escapeinside={&}{&},stepnumber=1,escapechar=\%]
    df = mesh_split(df, mesh, [0, 1])
\end{lstlisting}
it will additionally trigger the weight-update sharding optimization~\cite{xu2020automatic,rajbhandari2019zero}
where the weight will be unsharded on demand on the D dimension right before
this layer in the forward pass to reduce peak memory usage, and gradients will be communicated
with \texttt{ReduceScatter} instead of \texttt{AllReduce} in the backward pass and applied on
a sharded optimizer.

If the layer has a sparsely activated mixture-of-expert (MoE) architecture~\cite{shazeer2017outrageously}, it can
additionally have an expert E dimension in the \texttt{Einsum} equation on both operands and
the output, i.e., $EBD,EDF\to EBF$. To parallelize the experts across different devices, the
user only needs to annotate the E dimension on these tensors, e.g.,
\begin{lstlisting}[basicstyle=\footnotesize\selectfont\ttfamily\linespread{0.6},escapeinside={&}{&},stepnumber=1,escapechar=\%]
    ebd = mesh_split(ebd, mesh, [0, -1, -1])
    edf = mesh_split(edf, mesh, [0, -1, 1])
    ebf = mesh_split(ebf, mesh, [0, -1, 1])
\end{lstlisting}
In practice, the annotations on the activations \texttt{ebd} and \texttt{ebf} can be omitted and \sys
can infer them from the weights, unless the upstream or downstream layers have a different
pattern of parallelism.

\subsection{Pipeline parallelism reduced to tensor sharding}\label{sec:spmd-pipe}
Pipelining does not partition
individual operators or tensors, but partitions the graph into \emph{pipeline stages}. We
consider a constrained scenario of pipelining: all stages are the same subcomputation except
for having different weight values. This constraint is practical because a common way to
scale up models is to stack layers of the same pattern~\cite{gpipe,gpt32020}.

We reduce pipelining into a layer-wise sharding problem. Imagine that the layer computation is
rewritten in a stage-parallel way, where a leading layer/stage dimension L has been added to each tensor,
and the computation is entirely parallel on this dimension.
This transformation can be done by existing frontends' vectorization support like
JAX's \texttt{vmap()} and TensorFlow's \texttt{vectorized\_map()}.

\paragraph{Basic GPipe schedule.}
Pipelining requires data to be organized into \emph{microbatches}; each stage loops
over them one at a time~\cite{gpipe}. We use a \emph{shifting buffer} to pass data
across stages, and it also has a leading L dimension, as shown below.

\begin{lstlisting}[basicstyle=\footnotesize\selectfont\ttfamily\linespread{0.6},escapeinside={&}{&},stepnumber=1,escapechar=\%]
# Shifting buffer.
state = zeros([L, ...])
for i in range(num_microbatches + L - 1):
  # Shift state to the right by 1.
  from_prev_stage = pad_left(state, 1)[:-1]
  stage_ids = range(L)  # [0, 1, 2, ...]
  inp = next_input()
  input = elementwise_select(
    stage_ids == 0, inp, from_prev_stage)
  state = vmap(OneStageCompute)(input, ...)
\end{lstlisting}

The above is the Python code for the forward pass of vectorized pipelining.
During each iteration, the state buffer is shifted to
the right by one along L, so that the previous stage's last result is passed to
the next stage. The loop runs additional \texttt{stages - 1} iterations to wait for
the last stage to finish processing all the microbatches. The extra
iterations are equivalent to the \emph{bubbles} in earlier work~\cite{gpipe} that
describe the idle time due to data dependency, although the waiting devices
compute on padded data instead of being idle.

With the help of \texttt{vmap}
or \texttt{vectorized\_map}, this loop implementation can wrap a legacy single-stage
implementation \texttt{OneStageCompute} and turn it to a pipelined computation.

This user-level loop library does not implement distributed execution, and it can run on
a single device. To distribute it on multiple devices, users can simply annotate the
L dimension to be sharded, and \sys will turn the buffer shifting into cross-device
communication via \texttt{CollectivePermute}.

There are several benefits of this pipelining implementation. 1) It runs naturally
when combined with other types of parallelism in \sys, avoiding the need for extra
infrastructure. 2) It enables pipelining on part of the model, and switching to other
types of parallelism in other parts (Section~\ref{sec:conformer-perf}).
3) It benefits from the SPMD property, so that the infrastructure
is very simple and does not need to maintain
the interface between multiple programs.

\begin{figure}[t]
  \includegraphics[width=0.48\textwidth]{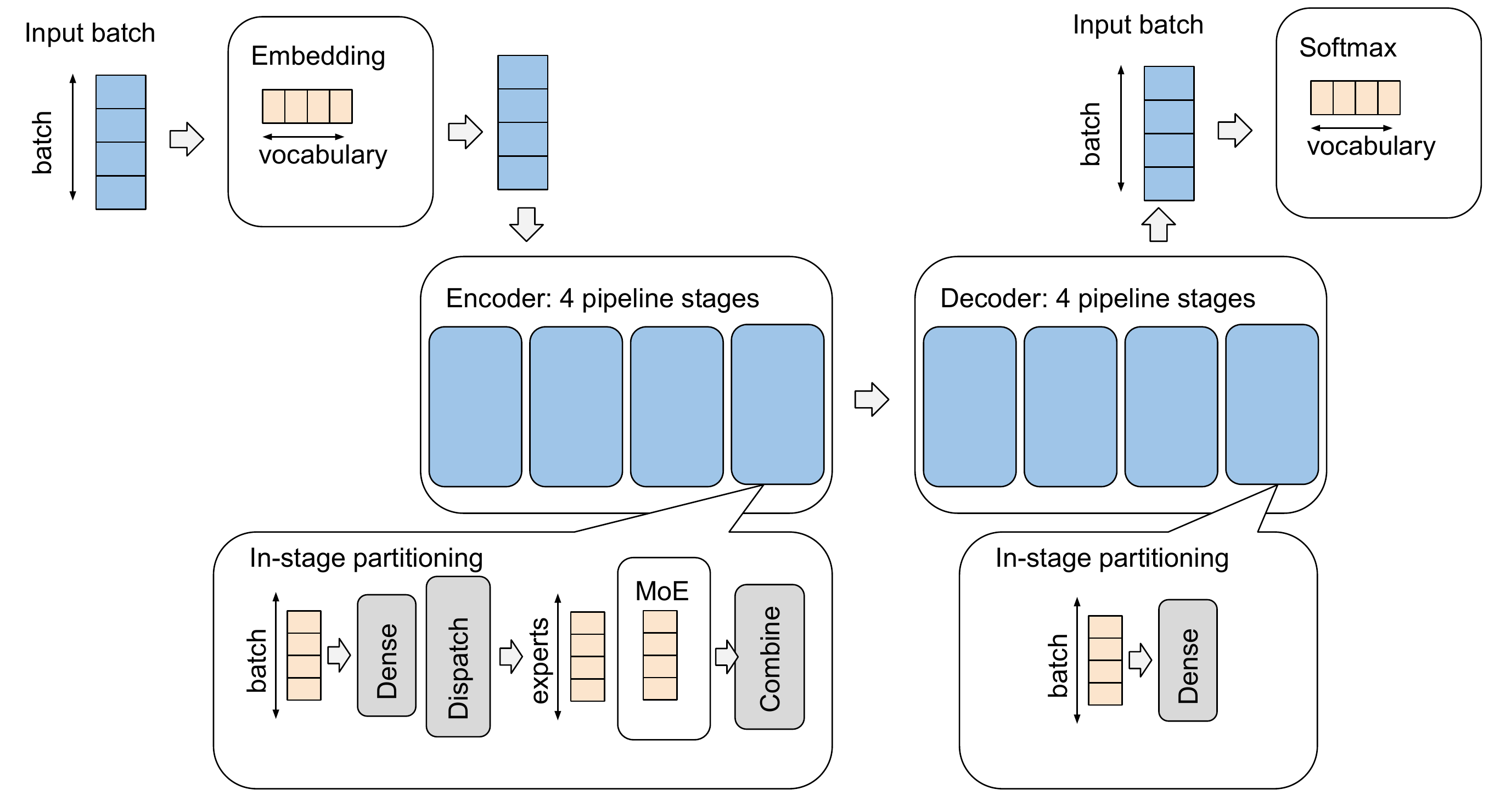}
\caption{A partitioning strategy over 16 devices organized as a logical 4x4 mesh for an encoder-decoder model, where the encoder conains MoE layers. Blue represents partitioning along the first mesh dimension X, and yellow represents partitioning along the second mesh dimension Y. X and Y are repurposed for different model components to achieve different parallelism modes. For example, the X dimension is used for data parallelism in the embedding and softmax layers, but used for pipeline parallelism in the encoder and decoder. The Y dimension is also used in different ways to partition the vocabulary, batch or model expert dimensions.}
\label{fig:multiple-pipes}
\end{figure}

It is limited to homogeneous pipeline stages, but this is not a constraint for
encoder-decoder models in general, since we can run separate pipelines for the encoder
and the decoder separately. Figure~\ref{fig:multiple-pipes} shows a configuration where
the encoder and decoder have their own pipelines that share the same set of devices, in
combination of sharding on other dimensions.

For heterogeneous stages that cannot be supported, we recommend integrating \sys with other pipeline
implementations~\cite{gpipe,terapipe,pipedream} and sharding each stage.

\paragraph{Circular schedule.}
This method also allows us to implement more advanced pipelinine scheduling algorithms.
For example, by assigning layers to devices in a non-contiguous manner
(e.g., Layers 0, 4, 8 to Device 0, Layers 1, 5, 9 to Device 1, ...), we can reduce the
bubble ratio with the same number of microbatches. It is implemented by adding an
extra dimension to represent the layers within a device.
We refer to this type of scheduling
as \textbf{circular pipelining}, which is similar to the interleaved schedule in \cite{narayanan2021efficient}.

\subsection{Manually partitioned subgraphs and pipelining}
\sys has a mechanism to allow power users to control exactly how a subgraph is partitioned,
by entering a \emph{manual partitioning} mode in the subgraph. Within this subgraph, the user
writes program with shard-sized shapes; outside the subgraph, the program is still to be
partitioned automatically by the compiler, and there are special conversion nodes to switch
between the modes. It was originally used to work around cases where
\sys was inefficient (see Section 3.2 in \cite{gshard-arxive}), but as the software matures
and advanced optimizations are added,
most of the use cases are no longer needed.

On the other hand, \sys pipelining (Section~\ref{sec:spmd-pipe}) becomes a popular use case
for manual-mode subgraphs, as an alternative to avoid \texttt{vectorized\_map()} in
TensorFlow since it supports only a subset of operators. Instead of doing
\texttt{vectorized\_map()} then partitioning the new stage dimension, we can simply
convert the inputs to manual mode before \texttt{OneStageCompute} thus removing
the stage dimension, and convert the outputs back to automatic mode.

To allow \sys to still partition other dimensions for data- or in-layer model-parallelism,
we extended the \textbf{manual mode to support subgroups} similar to partial replication, i.e.,
devices within a subgroup are manually partitioned, while devices across subgroups are
automatically partitioned. In this case, the sets of devices used as
pipeline stages are the manual subgroups.

\subsection{Intuitive sharding completion}
This section describes how \sys auto-completes the sharding on every tensor based on
limited user annotations. It is implemented as a compiler pass in XLA.

\paragraph{Preserved dimensions in operators.}
XLA operators typically preserve some dimensions from inputs to outputs. We simply
propagate sharding of such a dimension from inputs to outputs, and vice versa.
For example, a sharding annotation on the input batch dimension could be propagated
down to all layers' results (activations), and the same is true for image spatial
dimensions.

We decided to keep the sharding propagation simple, so it does
not try to create new sharding patterns on dimensions.
The propagation result may not always be optimal, but results will be
the most \textbf{intuitive} to users.
Users may insert more sharding annotations to instruct \sys to partition each tensor as desired. 

In comparison, a fully automatic approach could apply advanced algorithms to find the
best partitioning strategy (e.g., \cite{jia19}) beyond user annotations,
but there has not been a working implementation for our production need due to
different representations and incompleteness in problem formulation.
\sys instead focuses on propagating user intentions, though it may also be useful
in defining the search space for a fully automatic approach.

\begin{figure}[t]
  \includegraphics[width=0.4\textwidth]{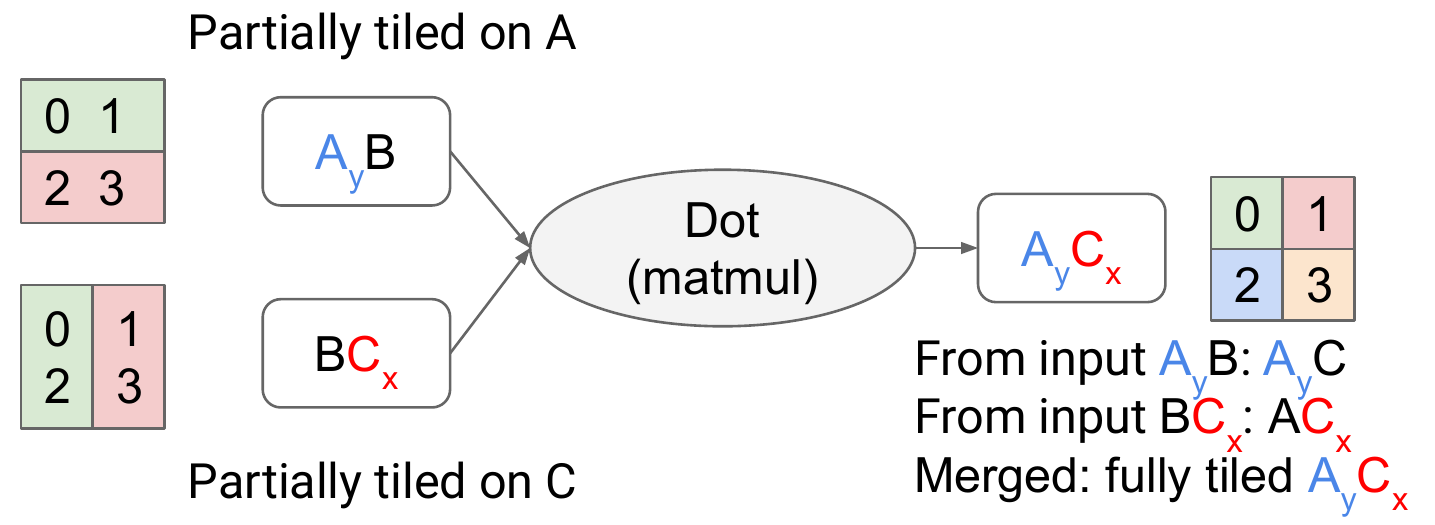}
\caption{Sharding propagation through a \texttt{Dot} operator. The result is merged
from both inputs. Different colors and subscripts of the letters represent sharding along different device mesh dimensions.}
\label{fig:dot-prop}
\end{figure}

\paragraph{Merging compatible shardings.}
The result of an XLA operator can inherit dimensions from different inputs. For
example, the \texttt{Dot} operator is a generalized matrix multiply, and
\sys infers sharding based on each operand;
\sys also tries to combine shardings from different operands if
they are compatible to form a more refined sharding.
A typical example of compatible shardings is two orthogonal partially tiled shardings
created by \texttt{mesh\_split} on different tensor dimensions, as shown in
Figure~\ref{fig:dot-prop}.

More formally, we use $Offset(S, d, i)$ to denote the \emph{shard offset} of device $d$ in
dimension $i$ according to sharding $S$. The shard offset describes the location of the
device's data partition within the original shape before partitioning. Then,
two shardings $S_0$ and $S_1$ are compatible with each other if there
exists a sharding $S$ where for each device $d$,
\begin{equation*}
Offset(S, d, i) == Offset(S_0, d, i)
\end{equation*}
 for every sharded dimension $i$ in $S_0$, and
\begin{equation*}
Offset(S, d, j) == Offset(S_1, d, j)
\end{equation*}
for every sharded dimension $j$ in $S_1$. In this case, $S$ is a merged sharding of $S_0$
and $S_1$.

Merging compatibile shardings is the key to support nested parallelism patterns with
simple user annotations. For example, if the user wants to mix data- and model-parallelism,
they could simply annotate the inputs to be sharded on the batch dimension along one device
mesh dimension, while
the layer weights are sharded on certain model dimensions along other mesh dimensions.
This way, the result of this
layer will be propagated with sharding on both dimensions.

\paragraph{Iterative, priority-based sharding propagation.}
To complete the sharding assignment on all tensors, \sys runs the propagation pass on the
entire graph in multiple iterations, and alternates between forward propagation (input to
output) and backward propagation (output to input). While it preserves initial user
annotations, shardings assigned by the pass could be refined incrementally over the
iterations. This means it changes the sharding on a tensor only when it finds a more
fine-grained sharding (possibily by combining with existing compatible sharding). This
property guarantees the pass could reach a fixed point after finite iterations.

\begin{figure}[h]
\begin{center}
\includegraphics[width=0.485\textwidth]{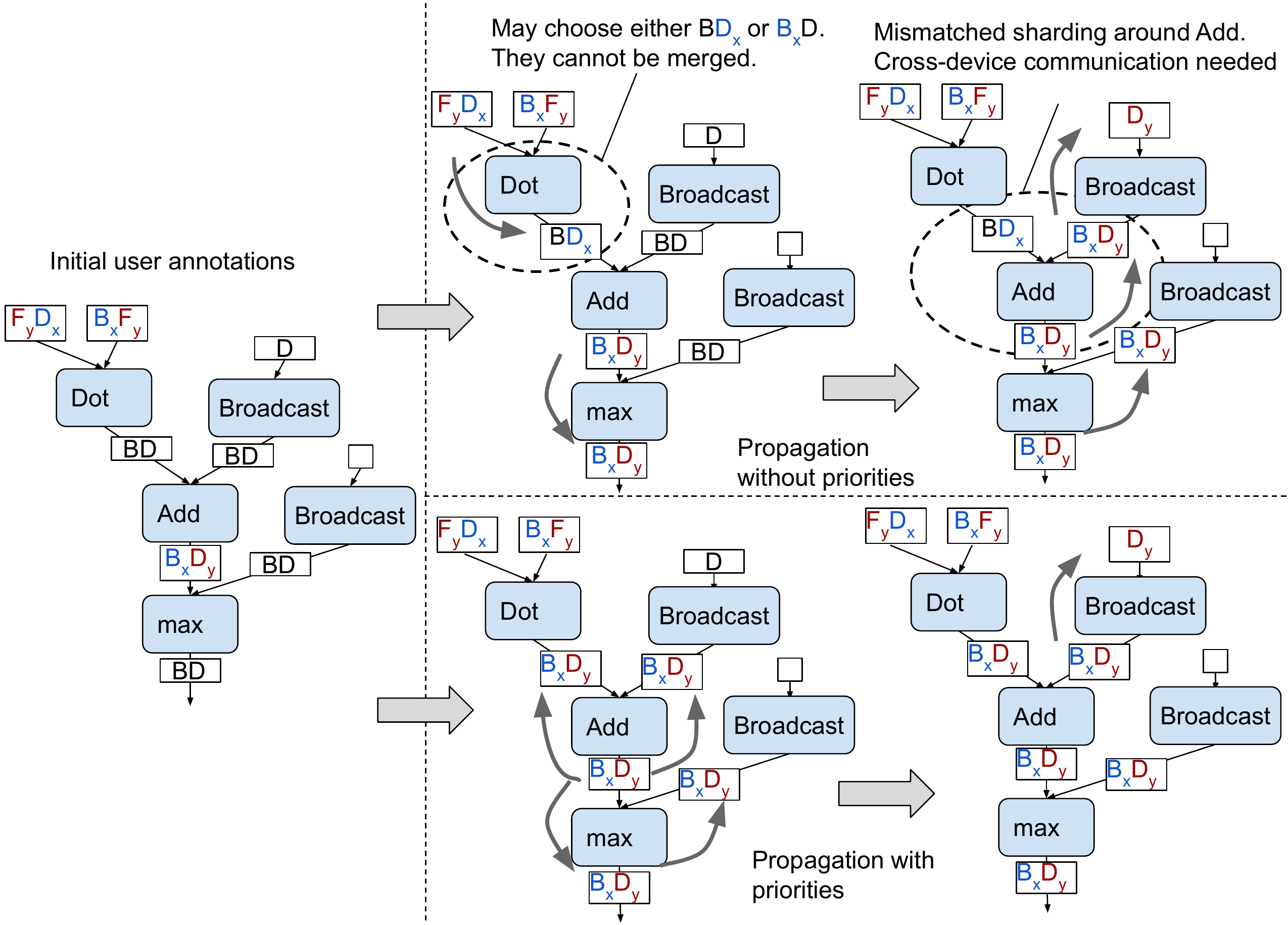}
\caption{\label{fig:priority_sharding_prop}Comparison between sharding propagation algorithms with and without priorities. The top-right figure shows a potential propagation process without priority, where tensors are visited in a topological order; since there are multiple ways to propagate through \texttt{Dot}, it may result in mismatched sharding specifications around an elementwise operator. The bottom-right figure shows the propagation process when elementwise operators are given a higher priority, where all the BD-shaped tensors are assigned the same sharding specification.}
\end{center}
\end{figure}

Figure~\ref{fig:priority_sharding_prop} illustration of the sharding propagation processes with and without priorities on a typical
linear layer followed by a ReLu activation function.

In practice, some use cases would require switching the sharding on a dimension in
different tensors (Section~\ref{sec:eval}), and the sharding
decision on each tensor will depend on the order in which the propagation pass
iterates over the operators. To produce the most intuitive sharding assignment,
\sys assigns a priority to the propagation through each operator from each direction.
Elementwise operators are assigned the highest priority to propagate
through in both directions, because 
there is no communication if their inputs/outputs are sharded consistently, and
propagating through elementwise operators
would be the most intuitive decision to users. Operators like
\texttt{Dot}, which add or remove dimensions are assigned lower priority. We also
assign different priorities to different directions; for example, the \texttt{Broadcast}
operator adds dimensions to the input by duplicating the data, so we assign higher
priority to the backward propagation, which helps to avoid potential communication
on the larger shape due to mismatched shardings.
See Figure~\ref{fig:priority_sharding_prop}.

\paragraph{Partial specification.}
By default, \sys does not change user-provided annotations. However,
in some cases the user wants to specify only a subset of tensor dimensions.
For example, the wrapper library for \sys pipeline (Section~\ref{sec:spmd-pipe})
wants to specify sharding only for the stage and the
\texttt{num\-\_microbatches} dimensions, while letting the wrapped layers determine
other dimensions of the variables and inputs. We could not use partial replication
to specify such cases in the default way, because that would prevent sharding
propagation from refining them. To support such cases, we extended
the annotation API to have a subset of \textbf{unspecified tensor dimensions}
subject to propagation changes.

\paragraph{Guide for users.}
Sharding propagation allows the users of \sys to skip manual annotations on
many intermediate results, such as those produced by layers like ReLu and
Batch Normalization. If users want explicit control over sharding decisions,
especially when tensors are sharded differently along a logical dimension,
they could focus on operators that significantly change the dimensions.
For example, if the inputs of a \texttt{dot} operator do not have compatible
shardings, there are multiple ways for the sharding propagation to infer the
output sharding; the user can explicitly annotate the output to precisely
control the sharding decision.

\subsection{API Integration in high-level frameworks}
\sys's sharding API is essentially an annotation on unpartitioned graphs. We
created a wrapper operator in TensorFlow, \texttt{XlaSharding}, to allow users
to instrument \sys by writing regular TensorFlow programs. From the user's point
of view, \texttt{XlaSharding} is semantically equivalent to an \texttt{Identity}
operator that passes the input through unchanged, but the sharding annotation can be specified
as an attribute. The TF2XLA bridge preserves the annotation when converting the
TensorFlow program to an XLA program.

Frameworks like TensorFlow also support automatic gradient calculation. 
It requires each operator to have a registered gradient
computation. We define the gradient of \texttt{XlaSharding} to be a copy of itself.
In this way, the backward computation will be annotated automatically with the same
sharding.
\section{The SPMD Partitioner}\label{sec:partitioner}

There are two options when implementing the partitioner: 1) creating a customized program
for each of the partitions (Multiple Programs Multiple Data, or MPMD), and 2) creating a
single program that works for all partitions (Single Program Multiple Data). We choose
SPMD because we aim to scale up to thousands of partitions, where compiling the many programs
would be prohibitively slow in MPMD. Compilation time is an important usability concern
because modern ML frameworks often include JIT optimizations and compilation, especially
for those targeting custom accelerators~\cite{xla,tvm,rotem2018glow}, and parallelizing the
compilation can be non-trivial because operators in different programs may need to be
globally scheduled to maintain correct communication order.

However, implementing the partitioner in SPMD creates unique challenges for production
ML compilers. This section covers the challenges for SPMD partitioning and the techniques
we use to solve them.

\subsection{Static constraints}\label{sec:static}
\paragraph{Static shapes.}
ML accelerators get their performance edge via specialized hardware, such as vector and
matrix units. In practice, the XLA compiler supports only limited degree of dynamism in
tensor shapes, in order to ease the development of highly efficient operator kernels.

\sys is designed to work even with full static shape constraints, but
static shapes create a challenge for SPMD partitioning. 
It's not always the case that all partitions have the same input/output shapes, because
dimensions may not be evenly divisible by the number of partitions. 
\sys
rounds up the size of the shape to a multiple of partition count, and the data in that
padded region can be arbitrary.

When creating certain partitioned operators, data in the padding area needs to be masked off.
For example, a \texttt{reduce} operator needs to prevent the padding data from affecting the
result, so \sys first replaces them with the identity value of the reduction.

The amount of padding can vary between different partitions. Although shapes are static,
several XLA operators accept dynamic offsets as operands, which allows \sys to express dynamic
padding area as a function of the partition ID.
Masking padded data can be expressed as a \texttt{Select} according to a mask calculated
by comparing \texttt{Iota} and an offset based on \texttt{PartitionId}.

\paragraph{Static operator configurations.}
XLA operators also have static configurations, like the padding, stride, and dilation defined in \texttt{Convolution}. However, different partitions may not execute with the same operator configuration. E.g., for a \texttt{Convolution}, the left-most partition applies padding to its left while the right-most partition applies padding to its right. 
The partitioner chooses a conservative configuration that makes some partitions to produce slightly more data than needed, then slices out the the irrelevant parts.

\subsection{Communication primitives}
Since the partitioner forces all the devices to run the same program, the communication patterns are regular. We use XLA's operators that provide MPI-style collective communications~\cite{mpi2.2}. 
\textbf{CollectivePermute} exchanges data among a list of source-destination pairs.
\textbf{AllGather} concatenates tensors from all participants following a specified order.
\textbf{AllReduce} performs elementwise reduction (e.g., summation) over the inputs from all participants.
\textbf{AllToAll} logically splits the input of each participant along one dimension, then sends each piece to a different participant. On receiving data pieces from others, each participant concatenates the pieces to produce its result.
\textbf{ReduceScatter} is semantically equivalent to an \texttt{AllReduce} followed by a \texttt{DynamicSlice} where each partition gets one slice of the fully reduced data. An efficient implementation has only half the cost of \texttt{AllReduce}.

\subsection{Halo exchange with dynamic bounds}

Certain operators have a communication pattern which involves partial data exchange with neighboring partitions, which we call \emph{halo exchange}. We use the \texttt{CollectivePermute} operator to exchange halo data between partitions.

\begin{figure*}[t!]
\subfloat[Convolution]{
 \includegraphics[width=0.28\textwidth]{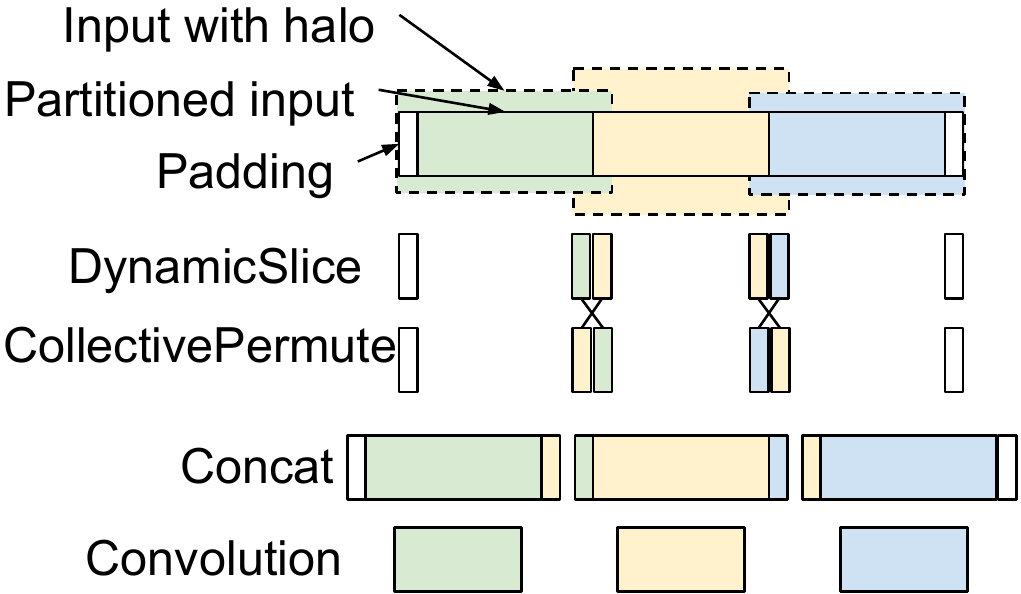}
 \label{fig:halo_exchange_conv}
}
\subfloat[Pad changes shard boundaries]{
 \includegraphics[width=0.28\textwidth]{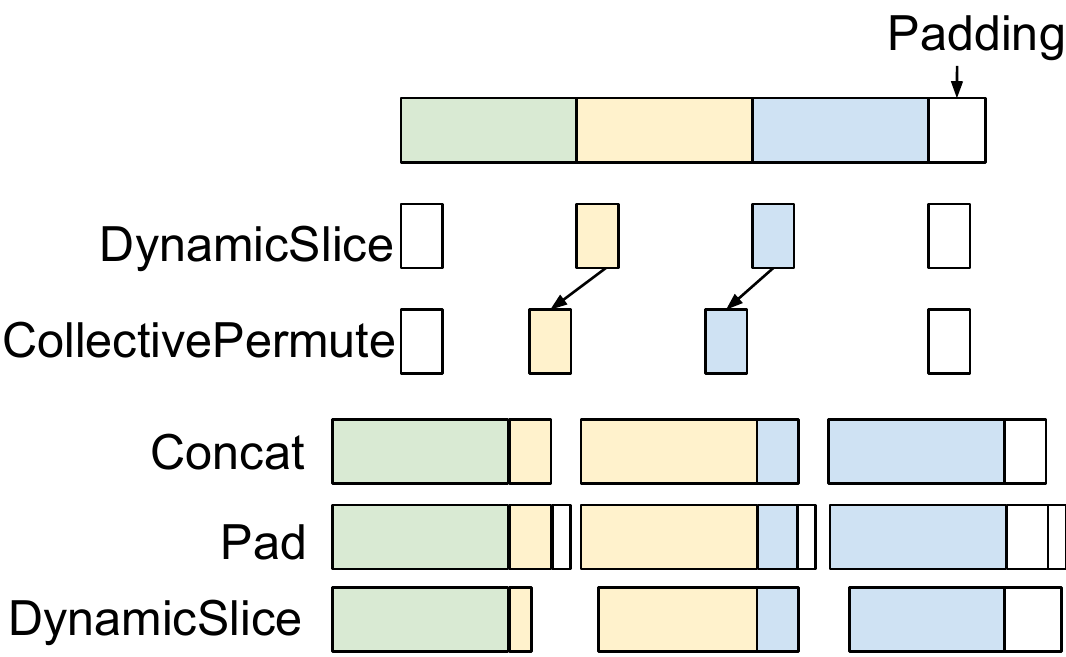}
 \label{fig:halo_exchange_pad}
}
\subfloat[Reshape changes uneven padding]{
 \includegraphics[width=0.28\textwidth]{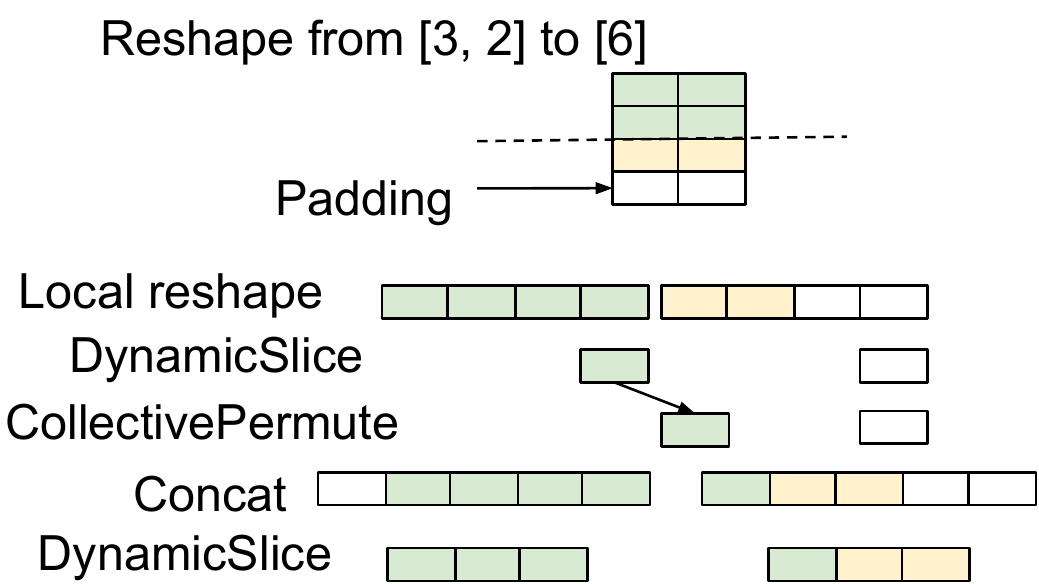}
 \label{fig:halo_exchange_reshape}
}
\caption{Halo exchange examples. Different colors represent data from different partitions.}
\end{figure*}

\paragraph{Windowed operators.} The most typical use case of halo exchange is for partitinoning window-based operators (e.g., \texttt{Convolution}, \texttt{ReduceWindow}), because neighboring partitions may require overlapping input data (Figure~\ref{fig:halo_exchange_conv}). In practice, halo-exchange for these operators often needs to be coupled with proper padding, slicing, and masking due to advanced use of window configurations (dilation, stride, and padding), as well as uneven halo sizes. See Section~\ref{sec:appendix-halo} in the Appendix for more details.

\paragraph{Non-constant halo size.} The amount of halo data needed by different partitions are often different. In such cases, \sys uses maximum halo size across partitions, then uses \texttt{Dynamic\-Slice} to remove excessive data in the halos. 
\sys supports the full set of configurations in the XLA \texttt{Convolution} operator, including arbitrary padding and dilation. These configurations add further complexity to the partitioner, but this can be ameliorated by applying careful padding and slicing.

\paragraph{Halo exchange for data formatting.}
Another use of halo exchange is for data formatting operators that change the size of the shape. For example, after a \texttt{Slice} or \texttt{Pad} operator, the shape of the tensor changes, and so do the boundaries between partitions. This requires us to realign the data on different partitions, which can be handled as a form of halo exchange (Figure~\ref{fig:halo_exchange_pad}).

Other data formatting operators may need halo exchange even when not changing the size, because the partitions may be uneven and the shapes are constrained to be static. For example, the \texttt{Reverse} operator reverses the order of elements in a tensor, but if it is partitioned unevenly, we need to shift data across partitions to keep the padding logically to the right of the result tensor. Another example is \texttt{Reshape}. Consider reshaping a tensor from (3, 2) to (6), where the input is unevenly partitioned in 2 ways on the first dimension (partition shape (2, 2)), and the output is also partitioned in 2 ways (partition shape (3)). There is padding on the input due to uneven partitioning, but after \texttt{Reshape}, the output tensor no longer has padding; as a result, halo exchange is required in a similar way to \texttt{Slice} (Figure~\ref{fig:halo_exchange_reshape}).

\subsection{Grouping and recursive partitioning}
\begin{figure}[ht!]
  \includegraphics[width=0.48\textwidth]{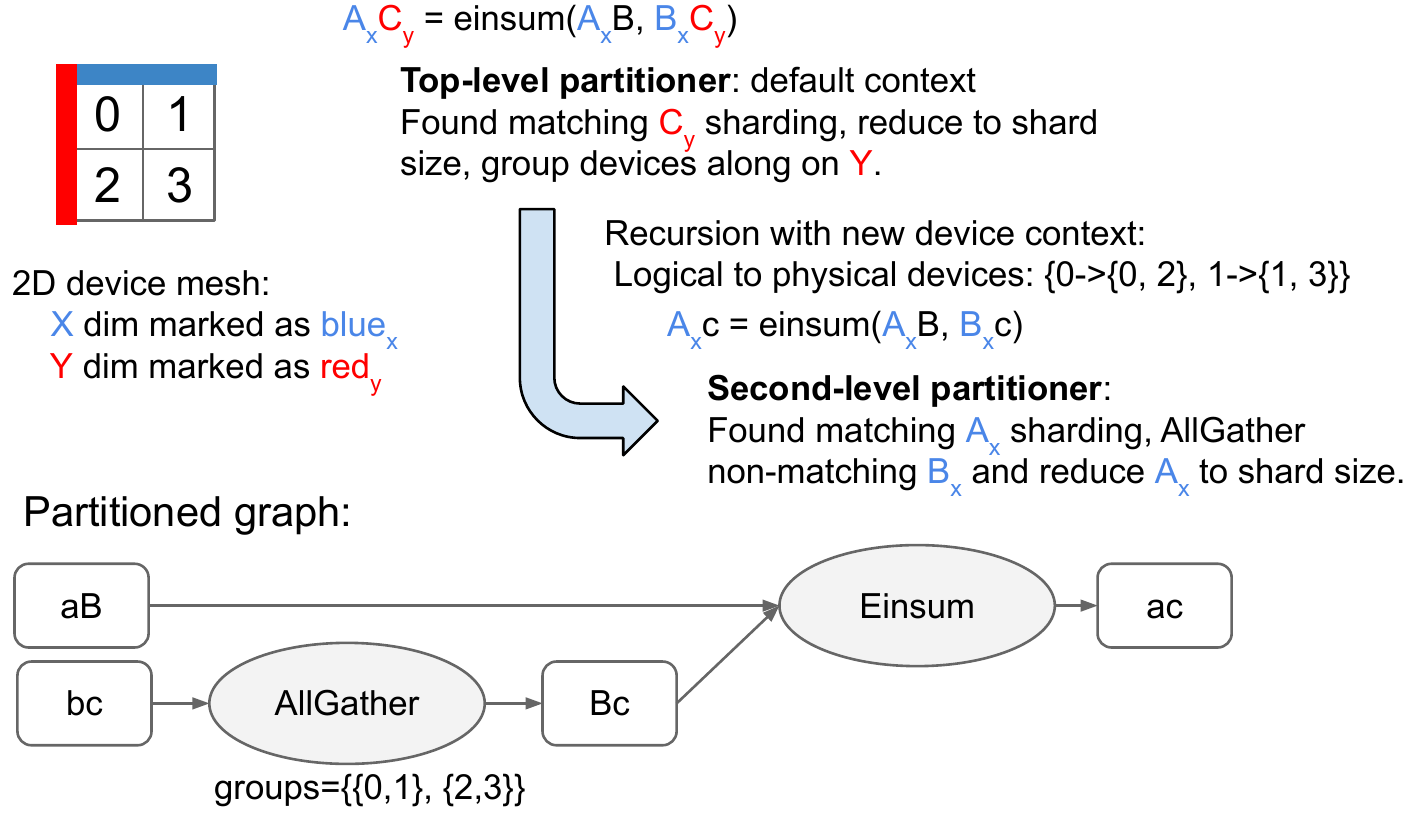}
\caption{Recursive partitioning of $AC = einsum(AB, BC)$. Letters in blue indicate dimensions sharded on the X dimension of the device mesh, while letters in red indicate dimensions sharded on the Y dimension. Lower case letters indicate sharded size. \texttt{AllGather} is created by the inner partitioner, where logical subgroups \{0,1\} are rewritten to \{\{0,1\}, \{2,3\}\} according to the context.}
\label{fig:recursive-partitioning}
\end{figure}

Many XLA and TensorFlow operators are rank polymorphic, where the same opcode can be used on tensors with
arbitrary number of dimensions; a classic example is the \texttt{Einsum} operator as we discussed in
Section~\ref{sec:api-examples}.  \sys's generic annotation API makes it convenient to combine
different parallelism modes by sharding multiple dimensions, so there is a need for the partitioner to
recognize not only fixed patterns but also nested cases.
In order to avoid manually writing partitioning rules on all combinations of patterns,
we developed a framework (Figure~\ref{fig:recursive-partitioning}) to recursively pattern-match each set of sharded dimensions, and partition nested cases.

First, we introduce a device context object for the partitioner, which defines how cross-partition collective operators are created based on given subgroups of devices, and how the partition ID is calculated. This context generalizes the concept of devices as virtualized \emph{logical partitions} via custom factory methods of collective operators and partition IDs.

Second, we introduce partition grouping. We can divide the devices into equally sized groups, and treat each group as a logical partition. Once we have such grouping, we can create a new partitioner context, where each logical partition is mapped to a group of devices. Within this context, whenever a collective operator is created, the logical partition IDs are rewritten to subgroups of original device IDs.

With this approach, \sys can perform recursive pattern matching on an \texttt{Einsum} op. For example, the partitioner detects whether there is a matching pattern on how the batch dimensions are partitioned in the inputs and the output. If such a batch-partitioned pattern exists, it could create a nested partitioner context by grouping the partitions across the batch dimensions and reducing the shape sizes; then it recursively calls the partitioner to handle other dimensions.

This technique greatly simplifies the partitioner's implementation of operators with rank polymorphism and complicated dimension configurations. For example, we applied it to \texttt{Convolution}, which allows the user to combine spatial partitioning and feature partitioning in the same operator.

\subsection{Resharding}
\sys always produces a valid partitioned graph regardless of what sharding annotations are provided. If the provided shardings are not the typical supported cases, the partitioner will perform \emph{resharding} on the inputs and/or outputs. Resharding can involve cross-device communications. It uses \texttt{AllGather} to replicate data along sharded dimensions, \texttt{AllToAll} to switch sharded dimensions, \texttt{CollectivePermute} to change device order, and \texttt{Dy\-namic\-Slice} to shard replicated dimensions.
\sys could use multiple steps of resharding to reach the desired sharding.




\subsection{Compiler optimizations for data formatting}
\paragraph{Pre-processing.}
Certain transformations may help produce faster programs simply by rearranging the data.
For example, a data rotation pattern \texttt{Concat(a[k:], a[:k])} that moves the first k elements to the end can
be recognized to avoid multiple rounds of halo exchanges in \texttt{Slice} and \texttt{Concat}. XLA does not define such an operator, but
we define a custom \texttt{SPMD\_Rotate} within the partitioner. Similar optimizations include merging a sequence of \texttt{Pad} and \texttt{Slice} operators, which is important to partition the pipelined program in Section~\ref{sec:spmd-pipe} where the shifting using \texttt{Pad} and \texttt{Slice} can be done with a single \texttt{CollectivePermute}.

\paragraph{Post-partitioning optimizations.}
The partitioner creates various data formatting operators in order to perform slicing, padding, concatenation, masking and halo exchange.
We leverage XLA's fusion capabilities, as well as new code motion optimizations for slicing and padding, to largely hide the overhead of data formatting. As a result, the run-time overhead is typically small.
\section{Case Study and Evaluation}\label{sec:eval}
This section demonstrates a few cases where we apply \sys to widely-used language, speech and image models.

We measure the model performance with \sys on the Cloud TPUv3 platform~\cite{tpu}, which has an XLA compiler backend so that it can execute the partitioned graphs produced by \sys. Each TPUv3 core has 16GB on-device memory, and the platform has high-speed homogeneous device-to-device links across many cores even if they are hosted on different machines, and these links form a 2D mesh. Such a platform is ideal to study the scalability of \sys, which achieves high compute utilization with operator sharding alone in many workloads. We also study pipeline parallelism  that works well in certain models and can be combined with operator sharding, which could be more useful for GPU platforms~\footnote{We have enabled \sys in XLA's GPU backend and verified its correctness, but do not have large-scale measurements for this paper.} where high-speed links typically exist only within a server host.

\subsection{Dense Transformer language model}
Transformer~\cite{transformer} is a widely-deployed model for tasks including translation, text generation and understanding. Its core consists of two alternating types of layers. Suppose the input to such layers is a tensor of shape (B, S, M), where B is the sample batch size, S is the sequence length per batch, and M is the model dimension for features.

The \emph{attention} layer can be described as:
\begin{equation*}
  y = \textrm{Attention}(W_Q \times x, W_K \times x, W_V \times x) \times W_O
\end{equation*}
where each of $W_Q$, $W_K$, $W_V$ is a weight matrix that projects $x$ into a tensor of shape (B, S, N, D). The N dimension is called the ``attention heads'', a parallel dimension during the computation of Attention(). The $W_O$ weight matrix projects the attention result back to shape (B, S, M).

The \emph{feed-forward layer} can be described as
\begin{equation*}
  y = \textrm{Relu}(W_{in} \times x) \times W_{out}
\end{equation*}
where $W_{in}$ is a weight matrix that projects $x$ into a tensor of shape (B, S, H), Relu() is elementwise, and the $W_{out}$ weight matrix projects the result back to shape (B, S, M). In the rest of the section, we refer to the combination of one attention layer and one feed-forward layers as one \emph{Transformer layer}.

Recent research has shown the benefit of scaling up the Transformer model~\cite{shazeer2018mesh,gpipe,gpt32020,lamda-blog,du2021glam} by increasing the number of layers and the sizes of dimensions M, H and N, which could reach hundreds of billions of parameters. We show that \sys allows these models to be trained efficiently by annotating just 7 tensors per Transformer layer (roughly 0.7\% of all tensors in the entire XLA graph). Internal attention computation and other layers like normalization do not need annotations, since \sys can find reasonable shardings automatically.

\begin{table*}[t]\footnotesize
    \centering
    \begin{tabular}{c|c|c|c|c|c|c|c|c?c|c}\toprule\hline
         &&  $W_Q,W_K,W_V$  & $W_O$ & $W_{in}$ & $W_{out}$ & \multicolumn{3}{c?}{Activations} & Memory usage & Communication \\\cline{1-9}
  \multirow{4}{*}{\rotatebox[origin=c]{90}{Shardings}}
  &Shape  &  MND & NDM & MH & HM & BSM & BSND & BSH & weight + activation & weight + activation \\\cline{2-11}
  &2D Attempt 1 & X,Y,\_ & Y,\_,X & X,Y & Y,X & \_,\_,X & \_,\_,Y,\_ & \_,\_,Y & O(1/(XY)) + (O(1/Y)+O(1/X)) & 0 + (O(1/Y)+O(1/X))  \\\cline{2-11}
  &2D Attempt 2 & X,Y,\_ & Y,\_,X & X,Y & Y,X & X,\_,\_ & X,\_,Y,\_ & X,\_,Y & O(1/(XY)) + O(1/X) & O(1/Y) + O(1/X) \\\cline{2-11}
  &2D finalized & X,Y,\_ & Y,\_,X & X,Y & Y,X & X,\_,Y & X,\_,Y,\_ & X,\_,Y & O(1/(XY)) + O(1/(XY)) & O(1/Y) + O(1/X) \\\cline{1-11}
  \bottomrule
    \end{tabular}
    \caption{Dense Transformer sharding annotations. X and Y are the two mesh dimensions.}
    \label{tab:transformer-sharding}
\end{table*}
 
\paragraph{2D sharding.}
One main goal of sharding is to make sure the model weights could fit into accelerator device memory.
We study a 2-dimensional sharding strategy that aims to scale to very large model sizes. We define the device mesh as a matrix of shape (X, Y), and annotate the tensors as specified in Table~\ref{tab:transformer-sharding}. We choose 2D mesh for 2 reasons: 1) it maps to the 2D topology of the TPU platform's device network, and 2) sharding a single dimension to a very small size would affect the compute efficiency on TPUs. We study 3 types of sharding configurations.

In a vanilla approach (2D Attempt 1 in Table~\ref{tab:transformer-sharding}), we shard H and N along Y, and M along X. The sharding annotations are consistent on all weight and activation tensors, avoiding any resharding. \sys adds subgrouped \texttt{AllReduce} to the graph.
However, 1) the activations are only partially sharded, so that activation memory during training becomes the bottlenneck for scalability; 2) the per-device weight is so small that it affects compute efficiency on TPUs.

Another approach (2D Attempt 2 in Table~\ref{tab:transformer-sharding}) is to use the same weight shardings, but switch the activations' X sharding to the batch dimension. Weights and activations have mismatching sharding along X, so \sys will perform subgrouped \texttt{AllGather} to unshard weights along X before each layer, but this avoids the need for \texttt{AllReduce} on BSH. This \texttt{AllGather} will not increase peak memory usage significantly since it is short-lived and the buffer will be freed as soon as the layer completes. In the backward pass, \sys will add a \texttt{ReduceScatter} on gradients due to batch and weight sharding on X. This behavior along X is conceptually the same as the weight-update/optimizer-state sharding technique~\cite{xu2020automatic,rajbhandari2019zero}. This approach solves the compute efficency problem, but it still suffers from an activation memory problem since the BSM tensor is still partially sharded.

\begin{figure}
\begin{center}
\includegraphics[width=0.48\textwidth]{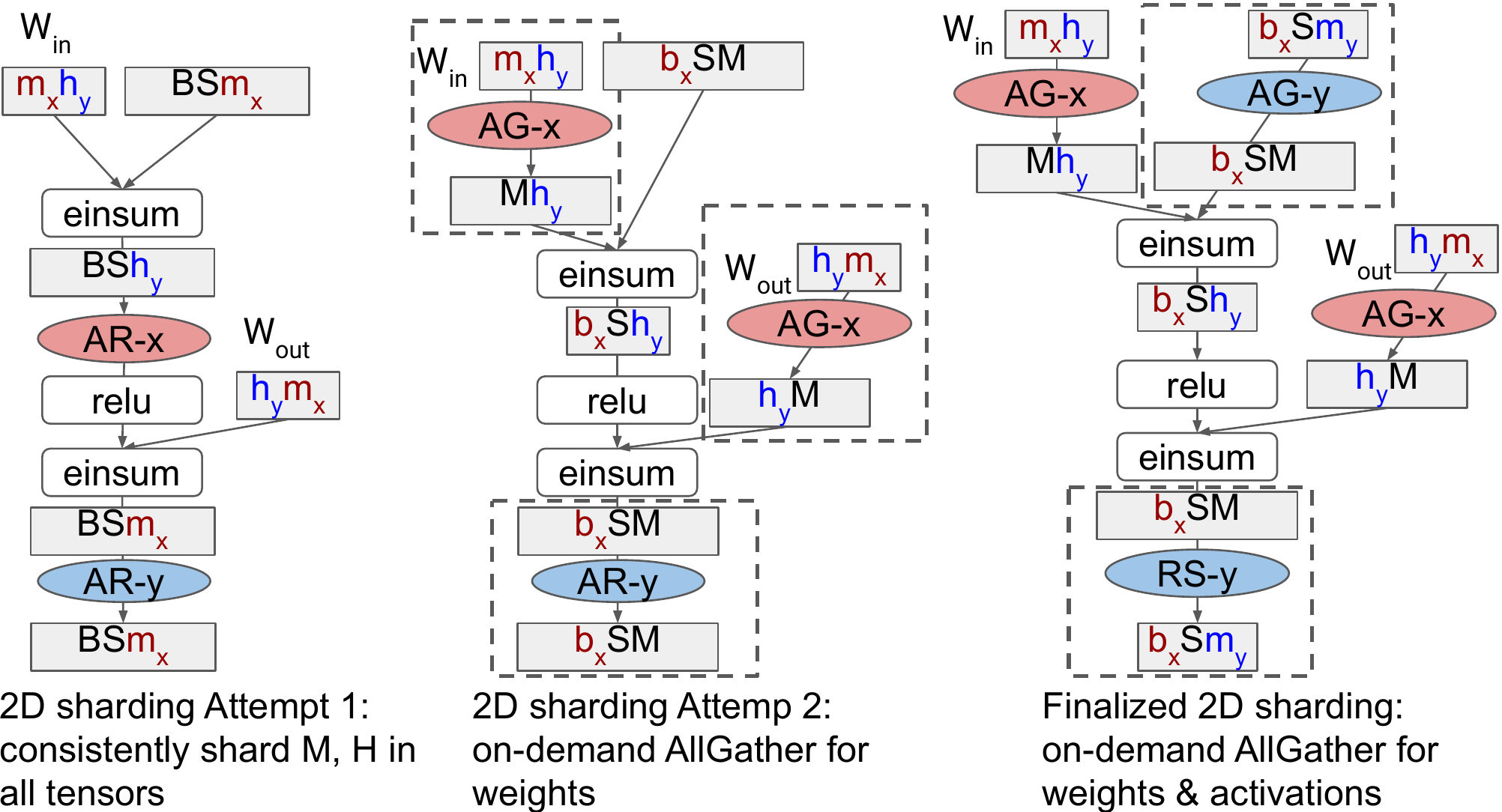}

\caption{Partitioned graphs for a Transformer feed-forward layer produced by \sys with the sharding annotations in Table~\ref{tab:transformer-sharding}. Lower-case letters indicate sharded dimensions, where different colors and subscripts denote different mesh dimensions. Collective operators: AR is \texttt{AllReduce}, AG is \texttt{AllGather}, and RS is \texttt{ReduceScatter}.}
\label{fig:partitioned-ffw}
\end{center}
\end{figure}

In our finalized sharding configurations (2D finalized in Table~\ref{tab:transformer-sharding}), we enhance Attempt 2 by further sharding the BSM activation's M dimension along Y. \sys will add a subgrouped \texttt{AllGather} for BSM to unshard M at the beginning of each layer, and replace the \texttt{AllReduce} on the output BSM with a \texttt{ReduceScatter}. The two new collective operators combined have comparable performance to the original \texttt{AllReduce}. In this approach, all long-lived tensors are fully sharded across all devices, so that peak memory can scale linearly when we increase the number of devices. We are now also able to use a relatively large batch size, which further helps TPU compute efficiency. This sharding configuration combines the benefit of data parallelism, weight-update sharding~\cite{xu2020automatic,rajbhandari2019zero} and in-layer model parallelism~\cite{shoeybi2019megatron}, and only requires 7 annotations in the model. Figure~\ref{fig:partitioned-ffw} shows the partitioned graphs for the 3 approaches.

\paragraph{Performance experiments.}
To evaluate the scalability of \sys, we measure the performance of training Transformer models that have many layers of large weights. We choose the following dimension sizes: M is 8192, H is 65536, N is 128, D is 256, and the vocabulary size is 32000. These dimensions are comparable to GPT-3~\cite{gpt32020}. We use a sequence length of 1024 for each input sample. Each Transformer layer (attention + feed-foward) has 2 billion parameters. We use 32-bit floating-point parameters, 16-bit floating-point activations, and Adafactor optimizer~\cite{adafactor}.

Scalability is evaluated by 2 sets of experiments: 1) a fixed model size (32 layers, 64B parameters) on different device topologies, and 2) different model sizes (32, 64, 128 and 256 layers) on a fixed device topology. Making the model deeper is more difficult than making it wider, because both compute and memory scale linearly with the model depth; in contrast, if we increase per-layer size to 4x by doubling the size of M, H and N, the activation memory only increases by 2x.

\begin{table}[t]\footnotesize
    \centering\setlength\tabcolsep{1pt}
    \begin{tabular}{c?c|c|c?c|c|c?c}\toprule\hline
    Parameter count & \multicolumn{3}{c?}{64B} & 128B & 256B & 512B & 1T\\\hline
    Layer count & \multicolumn{3}{c?}{32} & 64 & 128 & 256 & 128 \\\hline
    M dim & \multicolumn{6}{c?}{8192} & 16384 \\\hline
    H dim & \multicolumn{6}{c?}{65536} & 131072 \\\hline
    Total devices & 128 & 512 & 2048 & \multicolumn{4}{c}{2048} \\\hline
    Device mesh & (8,16) & (16,32) & (32,64) & \multicolumn{4}{c}{(32,64)} \\\hline
    Batch size & 64 & 256 & 1024 & 512 & 256 & 128 & 128 \\\hline
    Peak memory & 15.3GB & 13.3GB & 15.56GB & 14.0GB & 12.9GB & 13.6GB & 15.8GB \\\hline
    Step time   & 5.74s & 6.25s & 6.30s & 6.31s & 6.71s & 7.64s & 12.66s \\\hline
    FLOPS util & 62.7\% & 57.1\% & 56.9\% & 56.5\% & 54.1\% & 47.5\% & 55.6\% \\\hline
  \bottomrule
    \end{tabular}
    \caption{Benchmarks for dense Transformer with wide layers. The last column has 4x wider layers compared to others.}
    \label{tab:dense-bench}
\end{table}

Table~\ref{tab:dense-bench} shows the training performance of the above model configurations on the TPUv3 platform. \sys achieves close to linear scaling for such models, in terms of both memory and performance. For the 32-layer  model, when we increase the number of TPU devices by 2x, we can roughly double the maximum input batch size, while maintaining similar step time (within 10\%). On the same 2048-core device mesh, when the model depth increases by 2x, the maximum input batch size is roughly halved, while the step time increases only 7\% from 64B to 256B. When the model size reaches 512B, the batch size becomes small and the step time increases by 13\% over 256B. The last column shows a configuration with 1 trillion parameters, which is shallower than the 512B configuration, but each layer is 4x wider; as expected, the shallower 1T model can fit the same batch size as the 512B deeper model, while achieving higher efficiency due to wider layers. 
\sys achieves high overall FLOPS utilization (from 54\% to 62\% for most cases and 47.5\% for the 512B configuration) of the TPU devices.

\begin{table}[t]\footnotesize
    \centering\setlength\tabcolsep{4pt}
    \begin{tabular}{c?c|c|c?c|c|c}\toprule\hline
    Total devices & 64 & 128 & 256 & 64 & 128 & 256 \\\hline
    Device mesh & (4,16) & (8,16) & (8,32) & (16,4) & (16,8) & (32,8) \\\hline
    Batch size  & 48 & 96 & 192 & 48 & 96 & 192 \\\hline
    Peak memory & 12.4GB & 14.6GB & 14.8GB & 13.8GB & 12.7GB & 14.1GB \\\hline
    Step time & 3.56s & 3.43s & 5.71s & 3.10s & 3.37s & 3.36s \\\hline
    FLOPS util & 39.4\% & 40.5\% & 27.1\% & 45.7\% & 41.7\% & 41.3\% \\\hline
  \bottomrule
    \end{tabular}
    \caption{Benchmarks for 2D-sharded narrower dense Transformer. The model has 64 Transformer layers and 16 billion parameters.}
    \label{tab:narrow-bench}
\end{table}

\paragraph{Narrower dense Transformer.} Due to the relatively large batch size, activation communication is typically much more significant than weight/gradient communication. In Transformer layers, the amount of compute is $O(MH+MND)$, while the amount of activation communication is $O(M)$; therefore, with a fixed 2D device mesh, narrower models with smaller dimensions cannot utilize the TPU's compute power as well as wider models due to higher percentage of communication time. One mitigation is to reduce the Y dimension size of the mesh, while increasing the X dimension size accordingly.

We measure the performance of a 8x narrower model which is still too big to fit on a single device: M is 4096, H is 16384, N is 64, and D is 128. It has 64 layers and 16 billion parameters in total. The results are shown in Table~\ref{tab:narrow-bench}, which are consistent with our analysis above. While memory scaling is still roughly linear to the number of devices, having a smaller Y mesh dimension helps to achieve higher efficiency (with the same number of devices). With the same Y mesh dimension size, increasing the size of the X dimension allows to increase the batch size linearly with relatively constant step time.

\subsection{Combining pipelining and in-layer sharding}\label{sec:pipe-perf}
This section studies the performance of \sys pipelining described in Section~\ref{sec:spmd-pipe}. We choose the same narrower model in Table~\ref{tab:narrow-bench}, because activation communication is expensive in 2D-sharded narrower models if the Y mesh dimension is large, and another level of parallelism could be helpful.

We use a 3D device mesh (L, X, Y) where the leading L is used to shard pipeline stages. X and Y are used in a similar way to 2D sharding (Table~\ref{tab:transformer-sharding}), except that weights are not sharded along X; this is because pipelining requires the input to be divided into microbatches and weight sharding on X would incur expensive per-microbatch \texttt{AllGather}. Nonetheless, per-device weights are sufficiently small due to L sharding.

\begin{table}[t]\footnotesize
    \centering\setlength\tabcolsep{4pt}
    \begin{tabular}{c?c|c|c|c|c}\toprule\hline
    Device mesh & (2,16,8) & (4,16,4) & (4,16,4) & (8,16,2) & (8,8,4) \\\hline
    Pipeline stages & 2 & 4 & 4 & 8 & 8 \\\hline
    Batch size  & $16\times 64$ & $16\times 64$  & $32\times 32$  & $32\times 32$ &  $32\times 32$  \\\hline
    Peak memory & 14.9GB & 13.3GB & 13.1GB & 15.0GB & 11.5GB \\\hline
    Step time & 24.0s & 22.3s & 22.2s & 23.4s & 23.7s \\\hline
    Raw FLOPS util & 46.2\% & 58.0\% & 51.8\% & 54.8\% & 55.5\% \\\hline
    Bubbles & 5.6\% & 14.8\% & 8.0\% & 16.5\% & 16.1\% \\\hline
    Recompute & 22.3\% & 21.3\% & 21.7\% & 22.2\% & 20.6\% \\\hline
  \bottomrule
    \end{tabular}
    \caption{Benchmarks for pipelining on the same model in Table~\ref{tab:narrow-bench}. The device mesh has shape (L, X, Y), where L is used to shard pipeline stages, X is used for data parallelism, and Y is used for in-layer model parallelism. Raw FLOPS util is the reading from the profiler which counts padded compute (bubbles) as valid compute. The batch size is described as num\_microbatches $\times$ microbatch\_size.}
    \label{tab:narrow-pipe-bench}
\end{table}
We follow GPipe's approach~\cite{gpipe}, where a key difference from non-pipelined configurations is that we recompute forward pass intermediate results during the backward pass, a technique known as rematerialization~\cite{chen2016training} to lower peak memory usage. Rematerialization enables us to use larger number of microbatches to reduce pipeline bubbles.

Table~\ref{tab:narrow-pipe-bench} summarizes the performance results for configurations with 2, 4 and 8 stages. There are two major observations. 1) It is beneficial to balance the number of pipeline stages (L) and the number of in-layer model-parallel shards (Y), and the fastest configuration has 4 stages and 4 model-parallel shards. 2) Although these configurations have higher raw FLOPS utilization as reported by the profiler, the best one is still 24\% slower than 2D sharding with (X=32, Y=8) (Table~\ref{tab:narrow-bench}), because bubbles (compute on padded data) and recompute are overheads but counted as useful compute by the profiler.


\subsection{Pipelined Conformer models}\label{sec:conformer-perf}
Conformer~\cite{gulati2020conformer} is a speech model, and its backbone is stack of convolution-enhanced Transformer layers. It also has a convolutional subsampling layer before the backbone. We study scaling up this model by adding more layers, while each layer has dimensions M=3072, H=12288, N=16. It is even narrower than the one in Section~\ref{sec:pipe-perf}, and we find it is easier to scale using pipelining on the backbone, and switching to data parallelism on the layers before and after.  Table~\ref{tab:conformer-pipe-bench} shows the results of using \sys pipelining on two model configurations, with various batch sizes and both GPipe and circular schedules. The circular schedule is especially useful when the batch size is small, achieving similar bubble ratio compared to GPipe with much larger batch sizes. This pipelining approach has been used in BigSSL~\cite{zhang2021bigssl}.

\begin{table}[t]\footnotesize
    \centering\setlength\tabcolsep{2pt}
    \begin{tabular}{c?c|c|c?c|c|c}\toprule\hline
    Parameter count & \multicolumn{3}{c?}{6.47B} & \multicolumn{3}{c}{12.95B}  \\\hline
    Layer count & \multicolumn{3}{c?}{32} & \multicolumn{3}{c}{64} \\\hline
    Pipeline stages & \multicolumn{3}{c?}{8}  & \multicolumn{3}{c}{16} \\\hline
    Schedule & GPipe & GPipe & Circular & GPipe & GPipe & Circular \\\hline
    Batch size  & $64\times 1$ & $16\times 1$ & $16\times 1$  & $128\times 1$  & $32\times 1$ & $32\times 1$  \\\hline
    Peak memory & 13.8GB & 12.2GB & 14.1GB & 15.4GB & 12.8GB & 14.8GB \\\hline
    Step time & 8.40s & 2.80s & 2.30s & 18.37s & 5.58s & 4.42s  \\\hline
    Raw FLOPS util & 59.4\% & 57.9\% & 53.9\% & 60.5\% & 59.0\% & 50.6\% \\\hline
    Bubbles & 9.6\% & 29.9\% & 9.0\% & 10.4\% & 31.0\% & 10.0\% \\\hline
    Recompute & 22.9\% & 22.3\% & 21.3\% & 29.9\% & 22.8\% & 22.4\% \\\hline
  \bottomrule
    \end{tabular}
    \caption{Benchmarks for pipelining on Conformer models. The batch size is num\_microbatches $\times$ microbatch\_size. The circular schedule assigns layers to the stages in a round-robin manner.}
    \label{tab:conformer-pipe-bench}
\end{table}

\subsection{Sparse Transformer language model}
Mixture-of-experts (MoE) is a recently explored research direction in scaling up the Transformer model~\cite{shazeer2017outrageously,gshard-iclr,du2021glam}, where the feed-forward layer provides many independent model weights (called ``experts''), and each input token is operated on by only a couple of experts. Such models usually consist of both MoE and non-MoE layers in an alternating way.

\begin{figure}[t]
\subfloat[MoE layer partitioning]{
 \includegraphics[width=0.22\textwidth]{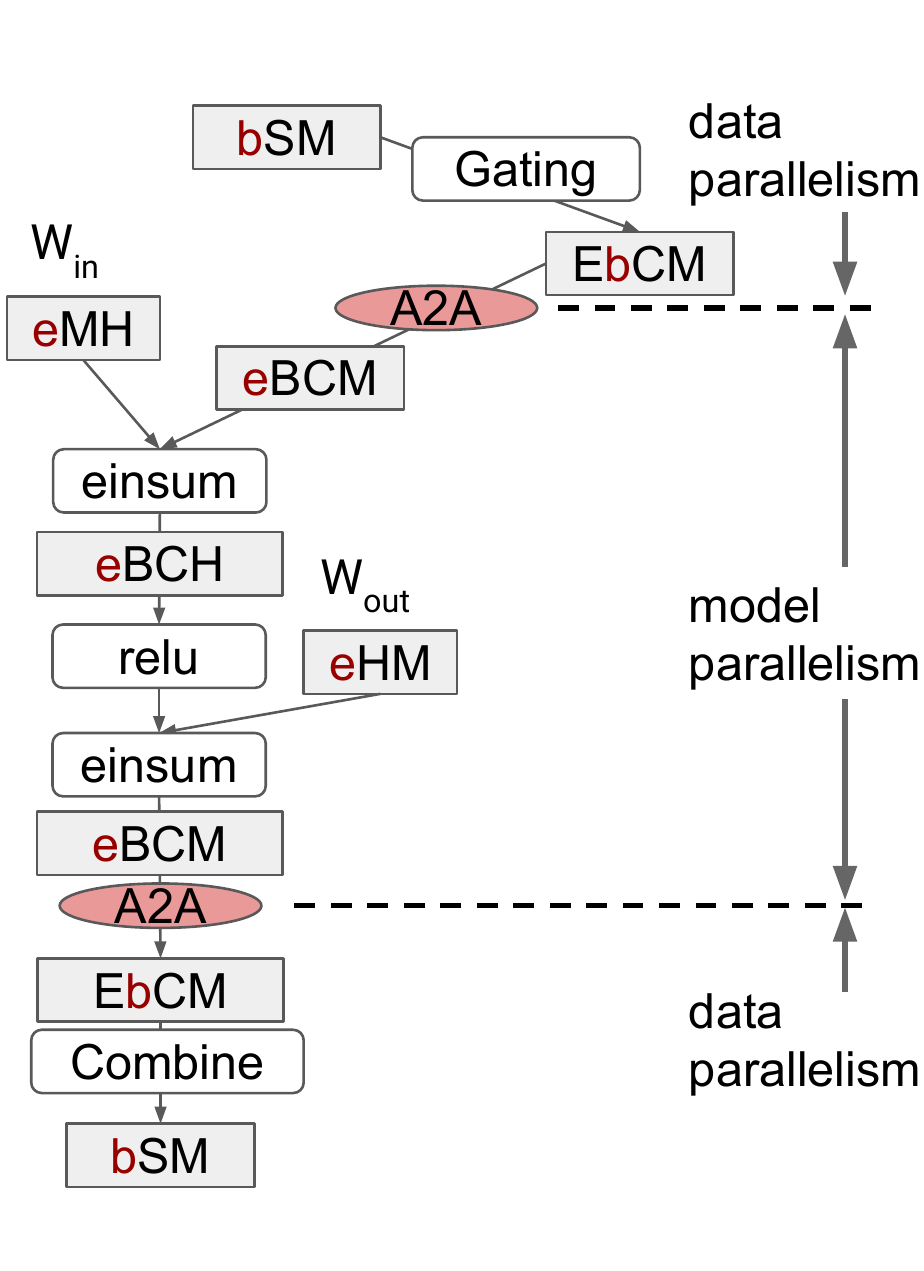}
 \label{fig:partitioned-moe}
}
\subfloat[Hybrid MoE layer partitioning]{
 \includegraphics[width=0.22\textwidth]{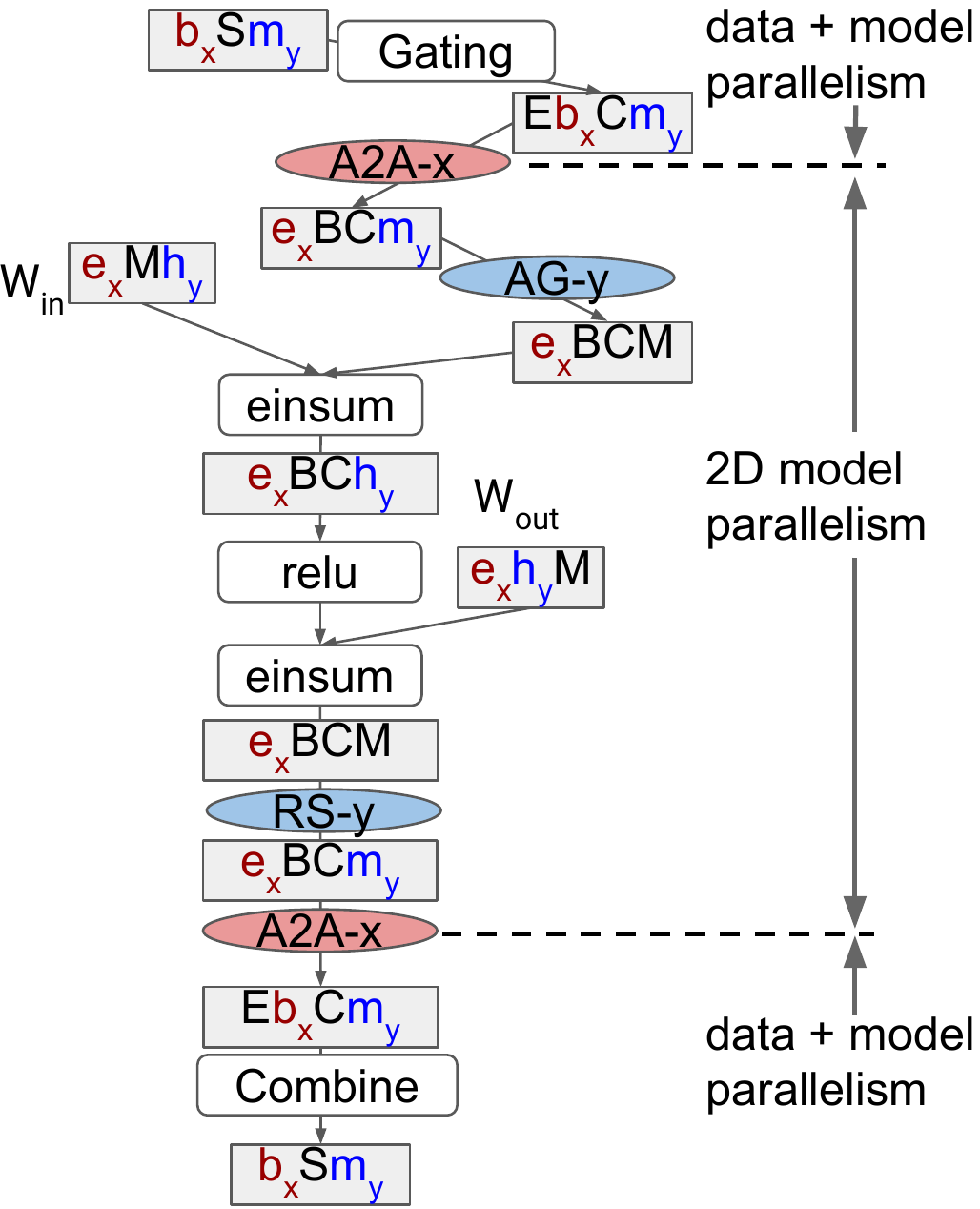}
 \label{fig:partitioned-hybrid-moe}
}
\caption{Partitioned graphs for MoE feed-forward layer and hybrid sparse and dense MoE layer. Lower-case letters indicate sharded dimensions, where different colors and subscripts denote different mesh dimensions. Collective operators: A2A is \texttt{AllToAll}, AG is \texttt{AllGather}, and RS is \texttt{ReduceScatter}.}
\end{figure}

The per-expert matrix multiplies in the MoE feed-forward layer can be expressed as \texttt{Einsum} operators with an extra parallel dimension, E (for experts), e.g., $EBCM,EMH\to EBCH$ and $EBCH,EHM\to EBCM$,  where C is the per-expert \emph{capacity}, i.e., the number of tokens to process per batch. Before each MoE layer, a \emph{gating} layer computes which experts each token goes to, transforming a (B, S, M) tensor to (E, B, C, M).

To scale up the model, we can increase the number of experts E, as well as the input batch size so that each expert still has a reasonable amount of data to process. We found a simple yet efficient way to partition such a model: use a 1D devicde mesh to shard the MoE layer's expert dimension, and switch to batch partitioning (data parallelism) in other layers.
For this type of sharding, \sys will insert \texttt{AllToAll} operators to switch the sharded dimension between E and B. In the backward pass, \sys inserts gradient \texttt{AllReduce} only for non-MoE layers because there is no data parallelism for MoE layers. Figure~\ref{fig:partitioned-moe} shows the partitioned forward pass.

\begin{table}[t]\footnotesize
    \centering
    \begin{tabular}{c?c|c|c|c}\toprule\hline
    Parameter count & 10B & 37B & 145B & 577B  \\\hline
    Experts per layer & 32 & 128 & 512 & 2048 \\\hline
    Device mesh & (32) & (128) & (512) & (2048) \\\hline
    Batch size & 128 & 512 & 2048 & 8192 \\\hline
    Peak memory & 10.8GB & 11.2GB & 11.2GB & 11.6GB \\\hline
    Step time   & 0.98s & 1.01s & 1.10s & 1.51s \\\hline
    FLOPS util & 58.2\% & 49.8\% & 49.8\% & 46.80\% \\\hline
    AllToAll time & 2\% & 6\% & 9\%  & 11\% \\\hline
  \bottomrule
    \end{tabular}
    \caption{Benchmarks for sparse MoE Transformer.}
    \label{tab:sparse-bench}
\end{table}

\paragraph{Performance experiments.} Except for the gating logic, the per-sample compute of this model is constant regardless of the number of experts. We study \sys's scalability on this model by fixing the size of each expert, while scaling the number of experts and the batch size. Table~\ref{tab:sparse-bench} shows the performance and memory results when we set the per-device expert count to 1 (could be any constant). The step time increases only from 0.98s to 1.10s from 32 to 512 experts, as expected; when the model further increases to 2048 experts, step time increases to 1.51s. The \texttt{AllToAll} communication time is roughly $O(\sqrt{\textrm{num\_devices}})$ on the 2D TPU device mesh, which contributes to the overhead with 2048 devices, but the main difference is that the gating compute becomes more significant with 2048 experts.

\subsection{Hybrid sparse and dense Transformer}
We consider a configuration of Transformer that combines the characteristics of sparse and dense models. There are still MoE layers, but each expert is sufficiently large such that we still need to shard each expert to fit the model weights. The largest model configuration, ``64B64E'', in GLaM~\cite{du2021glam} falls into this category. The non-MoE layers have the same size as a single expert in an MoE layer. We use a 2D device mesh of shape (X, Y), and the non-MoE layers are sharded the same way as the dense Transformer (Table~\ref{tab:transformer-sharding}). The MoE layers' H and N dimensions are sharded on Y, while the E dimension is sharded on X. The partitioned forward pass graph is shown in Figure~\ref{fig:partitioned-hybrid-moe}.

\begin{table}[t]\footnotesize
    \centering
    \begin{tabular}{c?c|c|c|c}\toprule\hline
    Parameter count & 33B & 57B & 420B & 804B \\\hline
    Experts per layer & 8 & 16 & 32 & 64 \\\hline
    H dim & \multicolumn{2}{c|}{32768} & \multicolumn{2}{c}{131072} \\\hline
    N dim & \multicolumn{2}{c|}{128} & \multicolumn{2}{c}{512} \\\hline
    Device mesh & (8,4) & (16,8) & (32,16) & (64,32) \\\hline
    Batch size & 32 & 128 & 128 & 512 \\\hline
    Peak memory & 12.3GB & 12.9GB & 11.9GB & 8.5GB \\\hline
    Step time   & 2.12s & 2.19s & 2.08s & 1.92s \\\hline
    FLOPS util & 55.3\% & 50.2\% & 50.8\% & 53.8\% \\\hline
  \bottomrule
    \end{tabular}
    \caption{Benchmarks for hybrid sparse/dense MoE Transformer.}
    \label{tab:hybrid-bench}
\end{table}

\paragraph{Performance experiments.} We scale the number of experts proportionally to X, and scale (batch size $\times$ per-expert weight size) proportionally to the total number of devices. Table~\ref{tab:hybrid-bench} shows the performance results of different configurations of the model using \sys. These configurations have roughly the same theoretical per-device compute and memory usage.  As expected, the measured step time and peak memory stay relatively constant as we scale the model; the variance is much smaller than pure sparse MoE configurations (Table~\ref{tab:sparse-bench}) because the hybrid configuration has fewer experts and much smaller \texttt{AllToAll} and gating overhead compared to the per-expert compute.

\subsection{Image spatial partitioning}
High-resolution images are often required in object detection and segmentation tasks, but they may not fit in the memory of a single device. This section discusses \emph{spatial partitioning} of convolution neural networks (CNNs). The goal is to shard activation tensors (along non-batch dimension) instead of model weights. As usual, \sys can express it with the same sharding API. In fact, sharding annotations are \emph{required only for the model inputs}; \sys can propagate the shardings on the spatial dimensions to all of the convolutional layers, because they share the same set of spatial dimensions. This technique has been used in MetNet-2~\cite{espeholt2021skillful}.

\begin{table}[t]\footnotesize
    \centering\setlength\tabcolsep{2pt}
    \begin{tabular}{c?c|c|c|c|c?c|c}\toprule\hline
    Device mesh & (1,1) & (1,2) & (1,4) & (1,8) & (1,16) & (4,16) & (2,32) \\\hline
    Image size & \multicolumn{5}{c?}{128x128x128} & \multicolumn{2}{c}{256x256x256} \\\hline
    Batch size & \multicolumn{5}{c?}{4} &  \multicolumn{2}{c}{8} \\\hline
    Peak memory & 14.3GB & 14.8GB & 7.9GB & 4.5GB & 2.7GB & 14.9GB & 8.52GB \\\hline
    Step time   & 2.99s & 1.56s & 0.79s & 0.39s & 0.19s & 1.66s & 0.76s \\\hline
    FLOPS util & 47.9\% & 43.7\% & 43.5\% & 43.5\% & 43.8\% & 20.5\% & 41.2\% \\\hline
  \bottomrule
    \end{tabular}
    \caption{Benchmarks for 3D U-Net with spatial partitioning. The first mesh dimension is used for data parallelism, and the second mesh dimension is used for spatial partitioning. Peak memory usage includes TPU-specific padding and does not scale linearly.}
    \label{tab:unet-bench}
\end{table}

\paragraph{Performance experiments.} We experimented with the 3D U-Net model~\cite{unet3d}, a popular dense 3D segmentation model which has been widely used in the medical image domain. With \sys, this model could run with the original resolution for CT images, which can be as large as 256x256x256. Spatial partitioning avoids downsampling which could affect accuracy. \sys makes this possible by simply annotating an input spatial dimension. We measure the performance of spatially partitioned 3D U-Net to evaluate \sys's convolution partitioning with halo exchange (Figure~\ref{fig:halo_exchange_conv}).

Table~\ref{tab:unet-bench} shows two sets of measurements. We first measure \sys's performance scaling with spatial partitioning only.
For image size 128x128x128, \sys achieves nearly linear scaling with a 15.7x step time reduction on 16 partitions.

We then measure the step time when combining spatial partitioning with data parallelism to keep the same number of total devices. This time we choose image size 256x256x256, which does not fit in a single device even with per-device batch size 1. We compare the results of 16-way and 32-way spatial partitioning, and found 32-way partitioning achieves 2x higher FLOPS utilization because it enables a higher per-device batch size.
\section{Related Work}
\label{sec:related}

Because programming in a distributed heterogeneous environment is challenging, particularly for high-level practitioners, deep-learning frameworks do not force users to specify precisely how the distributed computation is done. For example, TensorFlow~\cite{tensorflow} has support for data parallelism, and basic model parallelism with graph partitioning by per-node device assignment. Mesh TensorFlow~\cite{shazeer2018mesh} helps the user to build large models with SPMD-style per-operator partitioning, by rewriting the computation in a Python library on top of TensorFlow; in comparison, \sys partitions the graph in the compiler based on lightweight annotations, without requiring the user to rewrite the model.

\sys is generalized from the back-end system used in GShard~\cite{gshard-arxive}. \sys expands the sharding representation with partial tiling and manual subgroups, and implements new techniques like priority sharding propagation, recursive partitioning and pipelining, making it more suitable for combined parallelism patterns and support much more use cases beyond MoE language models.

Tofu~\cite{tofu} is another parallelization system for large models, but it supports only limited partition strategies (e.g., ``partition-n-reduce``), while \sys supports partitioning all dimensions of complex operators like \texttt{Convolution}.

Pipelining algorithms~\cite{gpipe,pipedream,terapipe,fan2021dapple,tarnawski2020efficient} focus on one type of parallelism, while \sys can be used either to express similar ideas with the help of vectorization (Section~\ref{sec:spmd-pipe}), or to work in combination of these implementations by additionally partitioning each pipeline stage.  Some of these works also provide orthogonal pipeline scheduling techniques that could be used in \sys to reduce pipeline bubbles.

Zero~\cite{rajbhandari2019zero} presents a set of optimizations to reduce memory redundancy in parallel training devices, by partitioning weights, activations, and optimizer state separately. \sys is more general: it does not distinguish these tensors and dimensions, and those specific partitioning techniques can be supported by annotating the corresponding tensor dimensions with a uniform API.
Weight-update sharding~\cite{xu2020automatic} is another automatic parallelization transformation that achieves optimizer state sharding similar to Zero, and conceptually it can be viewed as a special case for \sys.

Combination of in-layer model parallelism and pipelining has also been studied in previous works~\cite{deepspeed,narayanan2021efficient}. \sys provides a general implementation of many of their partitioning techniques under the same sharding annotation API. For example, the scatter/gather optimization across pipeline stages in \cite{narayanan2021efficient} is automatically enabled for all of the configurations in Table~\ref{tab:narrow-pipe-bench}, because the activations are fully sharded (scatter phase) and then combined on-demand (gather phase) in the next stage (\texttt{bSm} and \texttt{bSM} tensors in Figure~\ref{fig:partitioned-ffw}).

FlexFlow~\cite{jia2019beyond} uses automated search to discover the partitioning strategy of operators in a graph for better performance. While FlexFlow focuses on determining the partitioning policy, \sys focuses on the mechanisms to transform an annotated graph. The two are complementary to each other: \sys can be used to define a search space and perform the transformation, and automated search combined with \sys could provide a fully automated system.


\section{Conclusion}\label{sec:conclusion}

\sys is a largely automated parallelization system for machine learning computations. It offers a simple yet powerful API which is general enough to combine different typical parallelism patterns. \sys offers an intuitive auto-completion feature that enables the user to annotate only a few tensors to partition the entire model efficiently. We have demonstrated that \sys is able to partition several image, speech and language models on up to thousands of Cloud TPUv3 cores, with good and predictable performance and memory scaling.



\bibliography{main}

\begin{thebibliography}{10}

\bibitem{lamda-blog}
{LaMDA}: our breakthrough conversation technology.
\newblock \url{https://blog.google/technology/ai/lamda/}.

\bibitem{xla-semantics}
{XLA} operation semantics.
\newblock \url{https://www.tensorflow.org/xla/operation_semantics}.
\newblock Online; accessed 17 April 2021.

\bibitem{xla}
{XLA: Optimizing Compiler for TensorFlow}.
\newblock \url{https://www.tensorflow.org/xla}.
\newblock Online; accessed 17 April 2021.

\bibitem{deepspeed}
{DeepSpeed}: Extreme-scale model training for everyone.
\newblock
  \url{https://www.microsoft.com/en-us/research/blog/deepspeed-extreme-scale-model-training-for-everyone/},
  2020.
\newblock Online; accessed 17 April 2021.

\bibitem{tensorflow}
{\sc Abadi, M., Barham, P., Chen, J., Chen, Z., Davis, A., Dean, J., Devin, M.,
  Ghemawat, S., Irving, G., Isard, M., Kudlur, M., Levenberg, J., Monga, R.,
  Moore, S., Murray, D.~G., Steiner, B., Tucker, P., Vasudevan, V., Warden, P.,
  Wicke, M., Yu, Y., and Zheng, X.}
\newblock {TensorFlow: A System for Large-Scale Machine Learning}.
\newblock In {\em {12th USENIX Symposium on Operating Systems Design and
  Implementation (OSDI)}\/} (Savannah, GA, Nov. 2016).

\bibitem{bezanson2017julia}
{\sc Bezanson, J., Edelman, A., Karpinski, S., and Shah, V.~B.}
\newblock Julia: A fresh approach to numerical computing.
\newblock {\em SIAM review 59}, 1 (2017), 65--98.

\bibitem{jax2018github}
{\sc Bradbury, J., Frostig, R., Hawkins, P., Johnson, M.~J., Leary, C.,
  Maclaurin, D., and Wanderman-Milne, S.}
\newblock {JAX}: composable transformations of {P}ython+{N}um{P}y programs.

\bibitem{gpt32020}
{\sc Brown, T.~B., Mann, B., Ryder, N., Subbiah, M., Kaplan, J., Dhariwal, P.,
  Neelakantan, A., Shyam, P., Sastry, G., Askell, A., et~al.}
\newblock Language models are few-shot learners.
\newblock {\em arXiv preprint arXiv:2005.14165\/} (2020).

\bibitem{tvm}
{\sc Chen, T., Moreau, T., Jiang, Z., Zheng, L., Yan, E., Cowan, M., Shen, H.,
  Wang, L., Hu, Y., Ceze, L., Guestrin, C., and Krishnamurthy, A.}
\newblock {TVM}: An automated end-to-end optimizing compiler for deep learning.
\newblock In {\em Proceedings of the 13th {USENIX} Conference on Operating
  Systems Design and Implementation\/} (USA, 2018), OSDI'18, {USENIX
  Association}, p.~579–594.

\bibitem{chen2016training}
{\sc Chen, T., Xu, B., Zhang, C., and Guestrin, C.}
\newblock Training deep nets with sublinear memory cost, 2016.

\bibitem{spatial-partitioning}
{\sc Cheng, Y., Lee, H., and Berghammer, T.}
\newblock {Train ML models on large images and 3D volumes with spatial
  partitioning on Cloud TPUs}.
\newblock
  \url{https://cloud.google.com/blog/products/ai-machine-learning/train-ml-models-on-large-images-and-3d-volumes-with-spatial-partitioning-on-cloud-tpus},
  2019.
\newblock Online; accessed 17 April 2021.

\bibitem{unet3d}
{\sc {\c{C}}i{\c{c}}ek, {\"O}., Abdulkadir, A., Lienkamp, S.~S., Brox, T., and
  Ronneberger, O.}
\newblock {3D U-Net}: Learning dense volumetric segmentation from sparse
  annotation.
\newblock In {\em Medical Image Computing and Computer-Assisted Intervention --
  MICCAI 2016\/} (Cham, 2016), S.~Ourselin, L.~Joskowicz, M.~R. Sabuncu,
  G.~Unal, and W.~Wells, Eds., Springer International Publishing, pp.~424--432.

\bibitem{du2021glam}
{\sc Du, N., Huang, Y., Dai, A.~M., Tong, S., Lepikhin, D., Xu, Y., Krikun, M.,
  Zhou, Y., Yu, A.~W., Firat, O., Zoph, B., Fedus, L., Bosma, M., Zhou, Z.,
  Wang, T., Wang, Y.~E., Webster, K., Pellat, M., Robinson, K.,
  Meier-Hellstern, K., Duke, T., Dixon, L., Zhang, K., Le, Q.~V., Wu, Y., Chen,
  Z., and Cui, C.}
\newblock Glam: Efficient scaling of language models with mixture-of-experts,
  2021.

\bibitem{espeholt2021skillful}
{\sc Espeholt, L., Agrawal, S., Sønderby, C., Kumar, M., Heek, J., Bromberg,
  C., Gazen, C., Hickey, J., Bell, A., and Kalchbrenner, N.}
\newblock Skillful twelve hour precipitation forecasts using large context
  neural networks, 2021.

\bibitem{fan2021dapple}
{\sc Fan, S., Rong, Y., Meng, C., Cao, Z., Wang, S., Zheng, Z., Wu, C., Long,
  G., Yang, J., Xia, L., Diao, L., Liu, X., and Lin, W.}
\newblock {DAPPLE}: A pipelined data parallel approach for training large
  models.
\newblock In {\em Proceedings of the 26th ACM SIGPLAN Symposium on Principles
  and Practice of Parallel Programming\/} (New York, NY, USA, 2021), PPoPP '21,
  Association for Computing Machinery, p.~431–445.

\bibitem{tpu}
{\sc {Google Cloud}}.
\newblock {Cloud TPU}.
\newblock \url{https://cloud.google.com/tpu/}.
\newblock Online; accessed 17 April 2021.

\bibitem{gulati2020conformer}
{\sc Gulati, A., Qin, J., Chiu, C.-C., Parmar, N., Zhang, Y., Yu, J., Han, W.,
  Wang, S., Zhang, Z., Wu, Y., and Pang, R.}
\newblock Conformer: Convolution-augmented transformer for speech recognition,
  2020.

\bibitem{gpipe}
{\sc Huang, Y., Cheng, Y., Chen, D., Lee, H., Ngiam, J., Le, Q.~V., and Chen,
  Z.}
\newblock Gpipe: Efficient training of giant neural networks using pipeline
  parallelism.
\newblock {\em CoRR abs/1811.06965\/} (2018).

\bibitem{jia19}
{\sc Jia, Z., Zaharia, M., and Aiken, A.}
\newblock {Beyond Data and Model Parallelism for Deep Neural Networks}.
\newblock In {\em Proceedings of the Conference on Systems and Machine Learning
  (SysML)\/} (Palo Alto, CA, 2019).

\bibitem{jia2019beyond}
{\sc Jia, Z., Zaharia, M., and Aiken, A.}
\newblock {Beyond Data and Model Parallelism for Deep Neural Networks}.
\newblock In {\em Proceedings of the Conference on Systems and Machine Learning
  (SysML)\/} (Palo Alto, CA, 2019).

\bibitem{adam}
{\sc Kingma, D.~P., and Ba, J.~L.}
\newblock {Adam: a Method for Stochastic Optimization}.
\newblock In {\em {International Conference on Learning Representations
  (ICLR)}\/} (San Diego, CA, May 2015).

\bibitem{krizhevsky2012imagenet}
{\sc Krizhevsky, A., Sutskever, I., and Hinton, G.~E.}
\newblock {ImageNet} classification with deep convolutional neural networks.
\newblock In {\em Advances in neural information processing systems\/} (2012),
  pp.~1097--1105.

\bibitem{gshard-arxive}
{\sc Lepikhin, D., Lee, H., Xu, Y., Chen, D., Firat, O., Huang, Y., Krikun, M.,
  Shazeer, N., and Chen, Z.}
\newblock {GShard}: Scaling giant models with conditional computation and
  automatic sharding.
\newblock {\em {CoRR} abs/2006.16668\/} (2020).

\bibitem{gshard-iclr}
{\sc Lepikhin, D., Lee, H., Xu, Y., Chen, D., Firat, O., Huang, Y., Krikun, M.,
  Shazeer, N., and Chen, Z.}
\newblock {GShard}: Scaling giant models with conditional computation and
  automatic sharding.
\newblock In {\em International Conference on Learning Representations\/}
  (2021).

\bibitem{terapipe}
{\sc Li, Z., Zhuang, S., Guo, S., Zhuo, D., Zhang, H., Song, D., and Stoica,
  I.}
\newblock {TeraPipe}: Token-level pipeline parallelism for training large-scale
  language models, 2021.

\bibitem{mpi2.2}
{\sc {MPI Forum}}.
\newblock {MPI: A Message-Passing Interface Standard. Version 2.2}, September
  4th 2009.
\newblock available at: \url{http://www.mpi-forum.org} (Dec. 2009).

\bibitem{pipedream}
{\sc Narayanan, D., Harlap, A., Phanishayee, A., Seshadri, V., Devanur, N.~R.,
  Ganger, G.~R., Gibbons, P.~B., and Zaharia, M.}
\newblock {PipeDream}: Generalized pipeline parallelism for dnn training.
\newblock In {\em Proceedings of the 27th ACM Symposium on Operating Systems
  Principles (SOSP)\/} (2019).

\bibitem{narayanan2021efficient}
{\sc Narayanan, D., Shoeybi, M., Casper, J., LeGresley, P., Patwary, M.,
  Korthikanti, V., Vainbrand, D., Kashinkunti, P., Bernauer, J., Catanzaro, B.,
  Phanishayee, A., and Zaharia, M.}
\newblock Efficient large-scale language model training on gpu clusters, 2021.

\bibitem{paszke2019pytorch}
{\sc Paszke, A., Gross, S., Massa, F., Lerer, A., Bradbury, J., Chanan, G.,
  Killeen, T., Lin, Z., Gimelshein, N., Antiga, L., et~al.}
\newblock {PyTorch}: An imperative style, high-performance deep learning
  library.
\newblock {\em Advances in Neural Information Processing Systems 32\/} (2019),
  8026--8037.

\bibitem{rajbhandari2019zero}
{\sc Rajbhandari, S., Rasley, J., Ruwase, O., and He, Y.}
\newblock {ZeRO}: Memory optimization towards training a trillion parameter
  models.
\newblock {\em arXiv preprint arXiv:1910.02054\/} (2019).

\bibitem{dalle}
{\sc Ramesh, A., Pavlov, M., Goh, G., Gray, S., Voss, C., Radford, A., Chen,
  M., and Sutskever, I.}
\newblock Zero-shot text-to-image generation.
\newblock {\em CoRR abs/2102.12092\/} (2021).

\bibitem{rotem2018glow}
{\sc Rotem, N., Fix, J., Abdulrasool, S., Catron, G., Deng, S., Dzhabarov, R.,
  Gibson, N., Hegeman, J., Lele, M., Levenstein, R., Montgomery, J., Maher, B.,
  Nadathur, S., Olesen, J., Park, J., Rakhov, A., Smelyanskiy, M., and Wang,
  M.}
\newblock Glow: Graph lowering compiler techniques for neural networks, 2018.

\bibitem{shazeer2018mesh}
{\sc Shazeer, N., Cheng, Y., Parmar, N., Tran, D., Vaswani, A., Koanantakool,
  P., Hawkins, P., Lee, H., Hong, M., Young, C., et~al.}
\newblock Mesh-tensorflow: Deep learning for supercomputers.
\newblock In {\em Advances in Neural Information Processing Systems\/} (2018),
  pp.~10414--10423.

\bibitem{shazeer2017outrageously}
{\sc Shazeer, N., Mirhoseini, A., Maziarz, K., Davis, A., Le, Q., Hinton, G.,
  and Dean, J.}
\newblock Outrageously large neural networks: The sparsely-gated
  mixture-of-experts layer.
\newblock {\em arXiv preprint arXiv:1701.06538\/} (2017).

\bibitem{adafactor}
{\sc Shazeer, N., and Stern, M.}
\newblock {Adafactor: Adaptive Learning Rates with Sublinear Memory Cost}.
\newblock {\em CoRR abs/1804.04235\/} (2018).

\bibitem{shoeybi2019megatron}
{\sc Shoeybi, M., Patwary, M., Puri, R., LeGresley, P., Casper, J., and
  Catanzaro, B.}
\newblock {Megatron-LM}: Training multi-billion parameter language models using
  {GPU} model parallelism.
\newblock {\em arXiv preprint arXiv:1909.08053\/} (2019).

\bibitem{tarnawski2020efficient}
{\sc Tarnawski, J., Phanishayee, A., Devanur, N.~R., Mahajan, D., and
  Paravecino, F.~N.}
\newblock Efficient algorithms for device placement of dnn graph operators,
  2020.

\bibitem{transformer}
{\sc Vaswani, A., Shazeer, N., Parmar, N., Uszkoreit, J., Jones, L., Gomez,
  A.~N., Kaiser, L., and Polosukhin, I.}
\newblock {Attention Is All You Need}.
\newblock In {\em Proceedings of the 31st Conference on Neural Information
  Processing Systems (NIPS)\/} (Long Beach, CA, 2017).

\bibitem{tofu}
{\sc Wang, M., Huang, C.-c., and Li, J.}
\newblock Supporting very large models using automatic dataflow graph
  partitioning.
\newblock In {\em Proceedings of the Fourteenth EuroSys Conference 2019\/}
  (2019), pp.~1--17.

\bibitem{xu2020automatic}
{\sc Xu, Y., Lee, H., Chen, D., Choi, H., Hechtman, B., and Wang, S.}
\newblock Automatic cross-replica sharding of weight update in data-parallel
  training, 2020.

\bibitem{zhang2021bigssl}
{\sc Zhang, Y., Park, D.~S., Han, W., Qin, J., Gulati, A., Shor, J., Jansen,
  A., Xu, Y., Huang, Y., Wang, S., Zhou, Z., Li, B., Ma, M., Chan, W., Yu, J.,
  Wang, Y., Cao, L., Sim, K.~C., Ramabhadran, B., Sainath, T.~N., Beaufays, F.,
  Chen, Z., Le, Q.~V., Chiu, C.-C., Pang, R., and Wu, Y.}
\newblock Bigssl: Exploring the frontier of large-scale semi-supervised
  learning for automatic speech recognition, 2021.

\end{thebibliography}
\bibliographystyle{acm}

\appendix
\section{Appendix}

\subsection{XLA operators for dynamism}

Although XLA requires static shapes, data access can have dynamic offsets based on run-time values. Table~\ref{tab:dynamic_ops} summarizes a few operators in XLA that \sys uses to support dynamic behaviors across different partitions.

\begin{table}[h]
\begin{center}
\begin{tabularx}{.46\textwidth}{X}
 \toprule
 \textbf{PartitionId ()} \\
 Returns the partition id (integer) of the current device running the program. \\
 \hline
 \textbf{DynamicSlice (Tensor[T] operand, Tensor[int] start\_indices, List[int] size\_indices)} \\
 Returns a slice (of shape size\_indices) from the operand tensor where the offset is given as a tensor. Used to allow each partition to select different regions of a tensor, where the dynamic offset is often calculated from PartitionId. \\
 \hline
 \textbf{DynamicUpdateSlice (Tensor[T] operand, Tensor[T] update, Tensor[int] start\_indices)} \\
 Returns a tensor that replaces a sub-region of the operand tensor with the given update tensor. start\_indices specifies the offset of the region to be updated. Used similar to DynamicSlice, but to update different regions of a tensor by each partition.\\
 \hline
 \textbf{Iota (List[int] dimensions, int64 iota\_dimension)} \\
 Creates an integer tensor of dimensions shape, with sequentially increasing values from 0 along the iota\_dimension (similar to std::iota(), but multi-dimensional). Used to create a mask of a partitioned tensor on each partition by comparing it against the un-partitioned dimension size. \\
 \hline
 \textbf{Select (Tensor[bool] pred, Tensor[T] on\_true, Tensor[T] on\_false)} \\
 Selects each element between two tensors based on the predicate tensor. All three tensors have the same shape. Used to mask out a region of a tensor (e.g., padded region due to uneven partitions), where the mask is often computed using Iota.\\
 \bottomrule
\end{tabularx}
\vspace{0.1in}
\caption{XLA operators used by \sys to handle non-uniform behavior between partitions.}
\label{tab:dynamic_ops}
\end{center}
\end{table}

\subsection{Halo exchange details}\label{sec:appendix-halo}
We first introduce the window configurations for operators like \texttt{Convolution} that \sys has to consider. Each spatial dimension in the convolution has the following set of configurations.
\begin{asparaitem}
    \item \textbf{Stride} is the distance (in number of elements) that the window moves to produce the next output element.
    \item \textbf{Low/high padding} is the number of elements padded to the low/high end of the dimension in LHS (base).
    \item \textbf{Base dilation} is the dilation factor of the LHS, i.e., one plus the number of elements padded between every element (excluding low/high padding). No base dilation means the value is set to 1. Base dilation is applied before low/high padding.
    \item \textbf{Window dilation} is one plus the number of elements padded between every element in the RHS (window).
\end{asparaitem}

\begin{figure*}[ht]
\begin{center}
\subfloat[Convolution halo size depends on shard offset. Input data required by each partition are indicated by dotted windows of a unique color.]{
\includegraphics[width=0.45\textwidth]{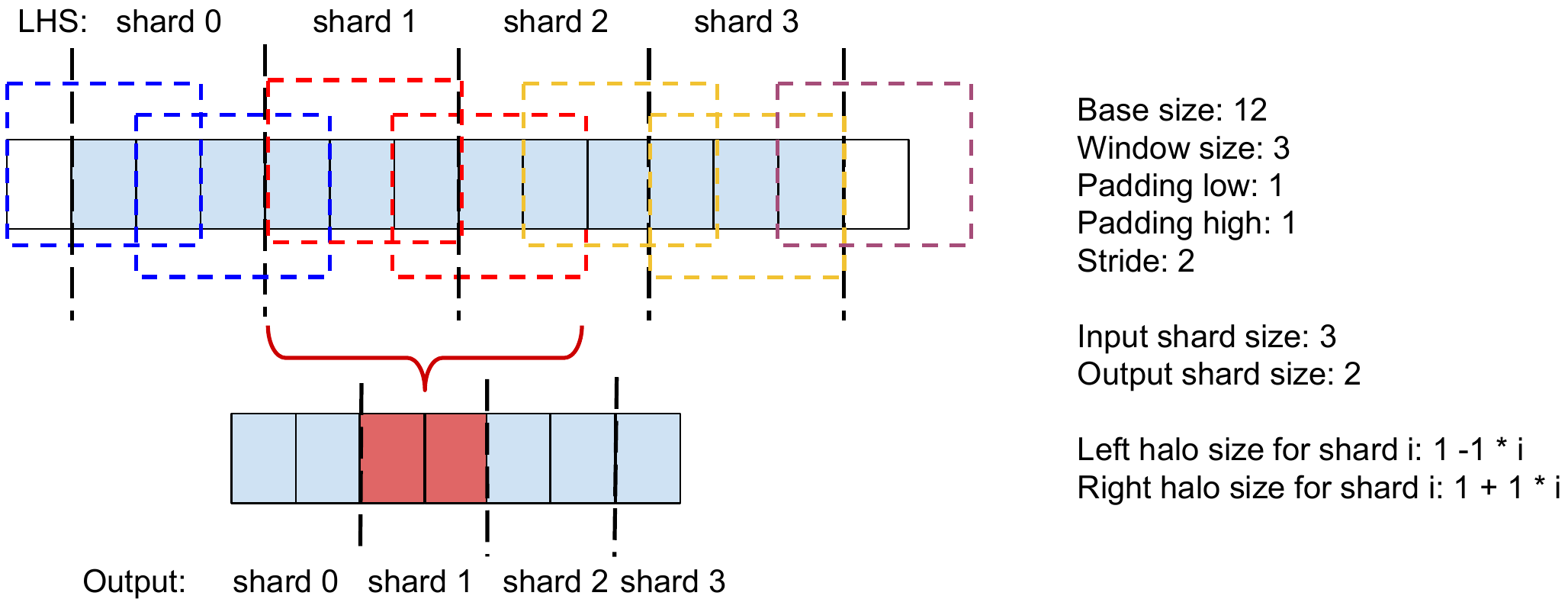}
\label{fig:conv_nonconstant_halo}
}
\subfloat[Sequence of operations for a general halo exchange.]{
\includegraphics[width=0.45\textwidth]{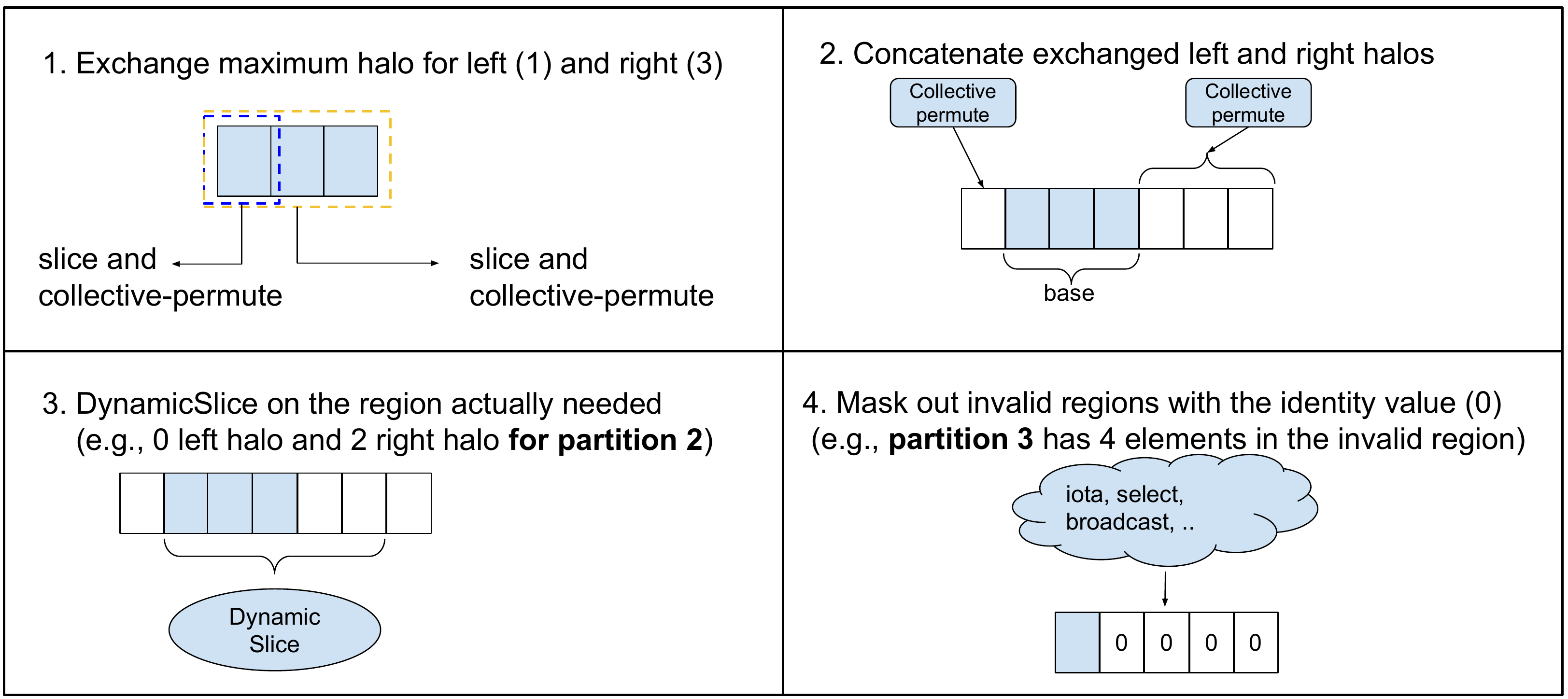}
\label{fig:conv_handling}
}
\caption{Non-constant halo size in a partitioned convolution and the solution with padding and slicing.}
\end{center}
\end{figure*}

\paragraph{Non-constant halo size.} We demonstrate that non-constant halo size is common using a simple example. Figure~\ref{fig:conv_nonconstant_halo} shows a 4-way partitioned convolution, where the right halo sizes for the partitions are (1, 2, 3, 4) and can be expressed as a linear function of the partition ID: \texttt{partition\_id + 1}.

Figure~\ref{fig:conv_handling} describes the sequence of operations for a general halo exchange. First, we calculate the maximum sizes of left and right halos across partitions and perform the halo exchange of the maximum size (Steps 1 and 2). Since some partitions may have excessive halos than needed, we use \texttt{DynamicSlice} (based on the partition ID) to get the valid region for the current partition (Step 3). Finally, some partitions may include garbage values (e.g., halos from out-of-range input data), so we apply masking as described in Section~\ref{sec:static}.

\begin{figure*}
\begin{center}
\includegraphics[width=0.75\textwidth]{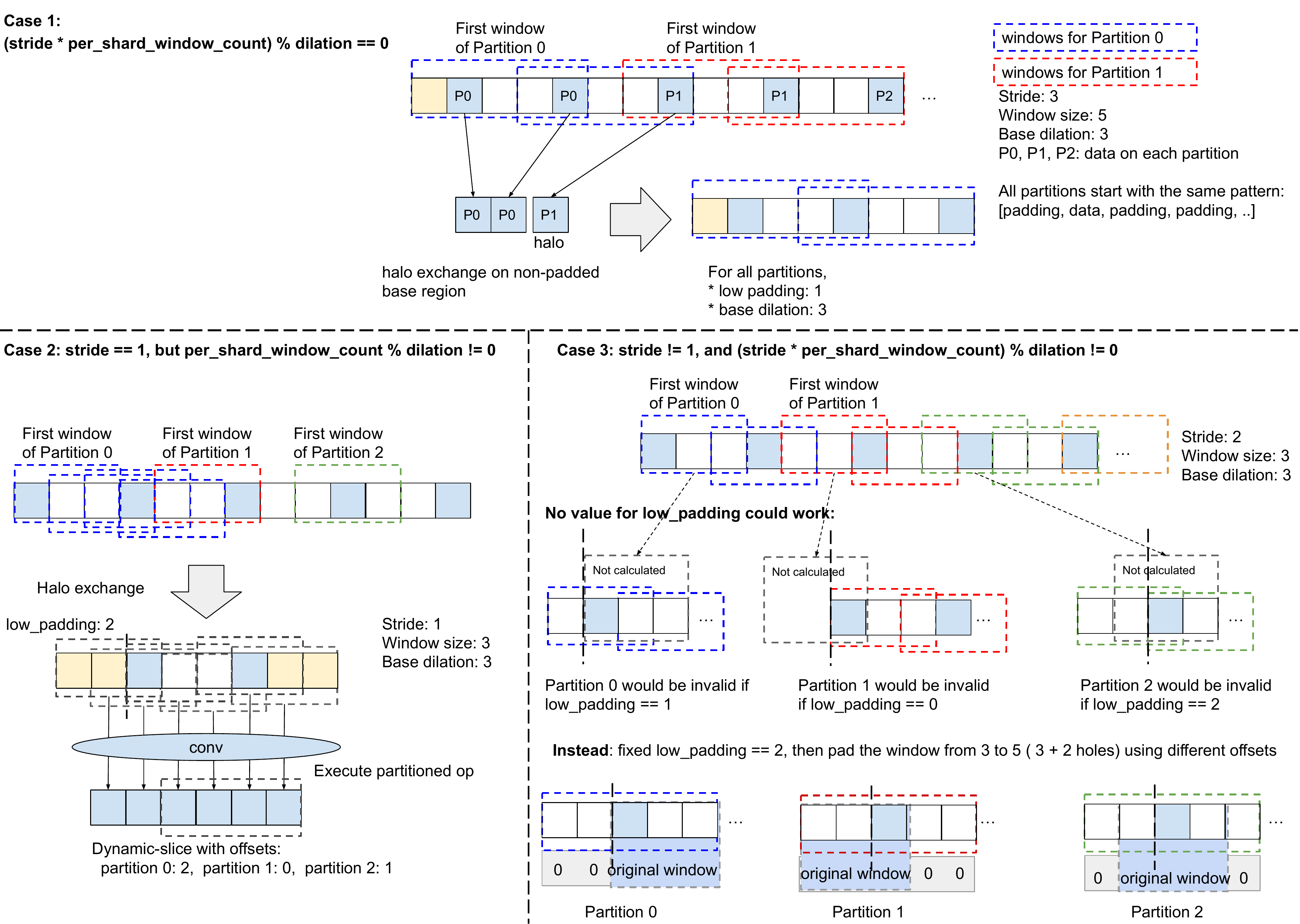}
\caption{\texttt{Convolution} partitioning with base dilation.}
\label{fig:base_dilation}
\end{center}
\end{figure*}

\paragraph{Base dilation.} Base dilation adds additional complexities to halo exchange, because the offset of each partition may be positioned at the dilation holes, which makes the edges have different behavior than the interior elements. We handle base dilation in 3 cases (Figure~\ref{fig:base_dilation}).

\begin{asparaitem}
    \item $stride \times per\_shard\_window\_count$ is divisible by $dilation$, where $per\_shard\_window\_count$ is the number of windows to be processed by each partition (i.e., the number of output elements for each partition). This condition guarantees that all partitions start with the same number of (interior or low) padding elements before the first data element in the LHS, so that we can use the same low padding configuration. Halo exchange occurs on the non-dilated and non-padded base region, and the limit index of required data for Partition $i$ can be represented as $a \times i + b$, where $a$ and $b$ are both integer constants. This limit determines the right halo size.

    \item $stride = 1$ but $per\_shard\_window\_count$ is not divisible by $dilation$. In this case, the low padding sizes vary on different partitions, but it is a static configuratio. Using \texttt{Pad} and \texttt{DynamicSlice} on the operand also would not work, because those operators would be applied before dilation, so everything would be multiplied by the dilation factor. Fortunately, with $stride = 1$, all positions on the padded and dilated base region are valid window starts, and we can use the maximum low padding on all partitions to ensure that each partition calculates all required windows, then perform a \texttt{DynamicSlice} on the output of the partitioned operator to remove unnecessary data. The limit index of required data on the non-padded base region for Partition $i$ can be represented as $(a \times i + b)/c$, where $a$, $b$ and $c$ are all integer constants and ``$/$'' is integer division.

    \item $stride\neq 1$ and $stride \times per\_shard\_window\_count$ is not divisible by $dilation$. If neither of the above conditions are true, different partitions could start with different number of padding elements, and not all offsets are valid window starts. Consider the last example in Figure~\ref{fig:base_dilation}. Whatever low padding we choose, some partition will be invalid, because valid windows could be skipped since $stride\neq  1$. A solution to this problem is to pad the window in addition to padding the base area. We can use the maximum low padding required by the partitions on the base area, and increase the window size by that low padding amount. The positions of the additional window padding vary on different partitions, which can be implemented as a \texttt{Pad} followed by a \texttt{DynamicSlice}. The window padding is used to mask off the unaligned elements in the base area, so that the start of the non-padding window element will be aligned with the desired start in the base area.
\end{asparaitem}

\paragraph{Window dilation.} If the RHS is replicated, window dilation only affects the effective window size when the operator is partitioned based on its LHS. If the dilated RHS is also partitioned, which typically occurs in the gradient computation of strided convolutions, handling window dilation is still simpler than handling base dilation, because there is no low/high padding on the RHS. We skip the implementation details.

\end{document}